\documentclass[prd,notitlepage,twocolumn,nofootinbib,amsmath,amssymb,longbibliography,eqsecnum]{revtex4-1}
\usepackage[utf8]{inputenc}
\usepackage[T1]{fontenc}
\usepackage{mathtools}
\usepackage{upgreek}
\usepackage{graphics}
\usepackage{graphicx}
\usepackage{dcolumn}
\usepackage{bm}
\usepackage{comment}
\usepackage{dsfont} 
\usepackage{amsmath,amssymb}
\usepackage{hyperref}
\usepackage{tabularx}
\usepackage{epstopdf}
\usepackage{tensor}
\usepackage{mathrsfs}
\usepackage[normalem]{ulem}
\usepackage[usenames]{color}
\hypersetup{
    colorlinks=true,
    linkcolor=blue,
    filecolor=magenta,      
    urlcolor=blue,
    citecolor=blue
}
\usepackage[dvipsnames]{xcolor}

\newcommand\be{\begin{equation}}
\newcommand\ba{\begin{eqnarray}}
\newcommand\ee{\end{equation}}
\newcommand\ea{\end{eqnarray}}
\newcommand\bw{\begin{widetext}}
\newcommand\ew{\end{widetext}}
\newcommand{\lb}{\left(}
\newcommand{\rb}{\right)}
\newcommand{\ETH}{\text{\DH}}

\newcommand{\UVA}{Department of Physics, University of Virginia, P.O.~Box 400714, Charlottesville, VA 22904-4714, USA}

\begin{document}

\title{Gravitational-wave memory effects in Brans-Dicke theory: Waveforms and effects in the post-Newtonian approximation}

 \author{Shammi Tahura}
 \affiliation{\UVA}

 \author{David A.~Nichols}
 \email{david.nichols@virginia.edu}
 \affiliation{\UVA}


 \author{Kent Yagi}
 \affiliation{\UVA}

\date{\today}

\begin{abstract} 
Gravitational-wave (GW) memory effects produce permanent shifts in the GW strain and its time integrals after the passage of a burst of GWs.
Their presence is closely tied to the symmetries of asymptotically flat spacetimes and the corresponding fluxes of conserved charges conjugate to these symmetries.
While the phenomenology of GW memory effects (particularly for compact-binary mergers) is now well understood in general relativity, it is less well understood in the many modifications to general relativity.
We recently, however, computed asymptotically flat solutions, symmetries, conserved quantities, and GW memory effects in one such modified theory with an additional scalar degree of freedom, Brans-Dicke theory.
In this paper, we apply our results from this earlier work to compute the GW memory effects from compact binaries in the post-Newtonian approximation.
In addition to taking the post-Newtonian limit of these effects, we work in the approximation that the energy and angular momentum losses through scalar radiation are small compared to the energy and angular momentum losses through (tensor) gravitational radiation.
We focus on the tensor (as opposed to scalar) GW memory effect, which we compute through Newtonian order, and the small differences induced by the radiation of scalar waves at this order.
Specifically, we compute the nonlinear parts of the tensor displacement and spin GW memory effects produced during the inspiral of quasicircular, nonprecessing binaries in Brans-Dicke theory. 
Because the energy radiated through the scalar dipole moment appears as a negative post-Newtonian order-effect, then in this approximation, the displacement memory has a logarithmic dependence on the post-Newtonian parameter and the spin memory has a relative minus-one-post-Newtonian-order correction; these corrections, however, are ultimately small because they are related to the total energy and angular momentum radiated in the scalar field, respectively.
At Newtonian order, the scalar radiation also gives rise to a sky pattern of the memory effect around an isolated source that differs from that of the memory effect in general relativity. 
\end{abstract}

\maketitle

\tableofcontents

\section{Introduction}

Following the first detection of gravitational waves (GWs) in 2015~\cite{Abbott:2016blz}, Advanced LIGO and Virgo have discovered almost 50 binary-merger events over their first two and a half observing runs~\cite{LIGOScientific:2018mvr,Abbott:2020niy}. 
These events allowed general relativity (GR) to be studied and tested in the dynamical and strong-field regime of the theory that had not been well constrained prior to LIGO and Virgo's observations.
The results of these tests of GR are summarized in a number of papers, for example~\cite{Abbott_IMRcon2,Yunes:2016jcc,TheLIGOScientific:2016pea,Abbott:2018lct,LIGOScientific:2019fpa,Berti:2018cxi,Berti:2018vdi}, which have all found the observed GWs to be consistent with the predictions of GR, within the statistical (and systematic) errors of the measurements. 
There remain GW effects that are still too weak to be extracted from the detector's noise, but which hold promise for revealing some of the more subtle nonlinear and dynamical effects in GR.
One such class of strong-field effects (which will be the subject of this paper) go by the name of gravitational-wave memory effects, because these effects are characterized by lasting changes in the GW strain and its time integrals that develop over the full history of the evolution of the system~\cite{Zeldovich:1974gvh,Christodoulou:1991cr,Blanchet:1992br}. 
Although these effects have not yet been observed, they could be detected in a population of binary-black-hole mergers measured by the Advanced LIGO and Virgo detectors over an observation period of several years~\cite{Lasky:2016knh,Hubner:2019sly,Boersma:2020gxx}; they could also be detected from an individual event by the space-based detector LISA~\cite{Favata:2009ii,Islo:2019qht}, third-generation ground-based detectors like Einstein Telescope or Cosmic Explorer~\cite{Johnson:2018xly}, or pulsar timing arrays~\cite{Wang:2014zls}.

In GR, gravitational-wave memory effects are closely related to the symmetry group of asymptotically flat spacetimes, the Bondi-Metzner-Sachs (BMS) group~\cite{Bondi:1962px,Sachs:1962wk,PhysRev.128.2851} and its generalizations.
We remind the reader that the BMS group is a semidirect product of an infinite-dimensional group of supertranslations (which are a kind of ``angle-dependent'' translations around an isolated source) and the Lorentz group. 
The charges conjugate to these symmetries are the relativistic angular momentum for the Lorentz symmetries [and angular momentum can be split into an intrinsic and a center-of-mass (CM) part, which correspond to the rotations and Lorentz boots symmetries, respectively] and the supermomentum for the supertranslations symmetries (see, e.g.,~\cite{Geroch:1977jn,Ashtekar:1981bq,Geroch:1981ut,Wald:1999wa}). 
There are more recent, and larger, symmetry algebras studied in which the Lorentz symmetries are extended to include all the conformal Killing vectors of the 2-sphere~\cite{Barnich:2009se,Barnich:2010eb,Barnich:2011mi} (not just the globally defined ones) or all smooth diffeomorphisms of the 2-sphere~\cite{Campiglia:2014yka,Campiglia:2015yka}. 
In the first case the symmetries are called super-rotations, and in the second, they are called super-Lorentz transformations; with the appropriate notion of supertranslations, they form the extended and generalized BMS algebras, respectively. 
The conserved charges corresponding to superrotations (and also the super Lorentz transformations) were called superspin and super CM when decomposed into the two parts of opposite parities~\cite{Flanagan:2015pxa,Nichols:2018qac,Nichols:2017rqr}.

As gravitational waves are radiated from a spacetime, fluxes and hence changes in the charges of the (generalized) BMS algebra generate GW memory effects. 
The displacement memory effect is produced by changes in the supermomentum charges, and the preferred shear-free frames before and after the burst of GWs are related by a supertranslation (see, e.g.,~\cite{Strominger:2014pwa,Flanagan:2015pxa}); the memory can be measured by nearby freely falling and comoving observers, who experience a lasting relative displacement. 
Memory effects associated with changes in the super-Lorentz charges and their fluxes are called the spin and center-of-mass (CM) GW memory effects~\cite{Pasterski:2015tva,Nichols:2017rqr,Nichols:2018qac}.\footnote{Note that the spin and CM GW memory effects are not related to a super-Lorentz transformation between certain canonical reference frames before and after a burst of GWs. 
There is a memory effect dubbed a refraction (or velocity-kick) memory that does correspond to a super-Lorentz transition between early time and late times~\cite{Compere:2018ylh}, but such solutions do not preserve the asymptotically flat boundary conditions of Bondi and Sachs~\cite{Bondi:1962px,Sachs:1962wk}.}
These memory effects are related to time integrals of the GW strain~\cite{Pasterski:2015tva,Nichols:2017rqr,Nichols:2018qac}, and they could be measured by nearby freely falling observers with an initial relative velocity~\cite{Flanagan:2018yzh}.
The (generalized) BMS flux balance laws provide a useful way to approximately compute the GW memory effects starting from gravitational waveforms without the GW memory effect, and this technique has been applied for the displacement, spin, and CM GW memory effects (see~\cite{Nichols:2017rqr,Nichols:2017rqr,Ashtekar:2019viz,Compere:2019gft,Mitman:2020bjf,Mitman:2020pbt,Khera:2020mcz,Liu:2021zys}). 

The extent to which the relationship between symmetries, conserved quantities, and memory effects might hold in modified theories of gravity is not immediately apparent. 
First, modified theories of gravity often admit additional polarizations of GWs~\cite{Will:2014kxa,Chatziioannou:2012rf,Zhang:2019iim}, and it is natural to suppose there would be memory effects associated with the additional polarizations (and there are such effects~\cite{Hou:2020tnd,Tahura:2020vsa,Du:2016hww,Koyama:2020vfc}).
Second, the asymptotic fall-off conditions on the fields in the modified gravity theory (i.e., what one defines as an asymptotically flat solution) could conceivably differ from those in GR.
Different boundary conditions could lead to different symmetry algebras and conserved quantities; this could in turn change both the types of possible memory effects and their relations to fluxes of conserved quantities.
It was with these considerations in mind that we recently investigated asymptotic symmetries, conserved quantities, and memory effects in one specific modified theory of gravity: Brans-Dicke theory~\cite{Brans:1961sx}. 

Brans-Dicke (BD) theory, is a so-called scalar-tensor theory in which a massless scalar field is coupled nonminimally to gravity (namely, there is a product of the scalar field and Ricci scalar in the theory's action). 
Scalar-tensor theories appear in the context of string theory and in phenomenological models used to explain the late-time acceleration of the Universe~\cite{Brax:2004qh,Baccigalupi:2000je,Riazuelo:2001mg} as well as cosmic inflation~\cite{Clifton:2011jh,Barrow:1990nv}. 
The massless scalar field leads to an additional ``breathing'' polarization of GWs in BD theory; for a freely falling ring of particles, this breathing mode produces a relative contraction and expansion in the plane transverse to the direction of GW propagation.

Recently, we (in~\cite{Tahura:2020vsa}) and Hou and Zhu (in~\cite{Hou:2020tnd}) independently studied BD theory in the Bondi-Sachs framework.
In addition to arriving at similar boundary conditions for asymptotically flat solutions in BD theory, we both showed that the symmetry group of asymptotically flat spacetimes in such a theory is the same as the (extended or generalized) BMS group.  
In~\cite{Tahura:2020vsa}, we observed that in addition to the GW memory effects present in GR, there are two more memory effects in BD theory related to the breathing-mode polarization.
We will adopt the nomenclature used in~\cite{Du:2016hww}, which calls the GW memory effects in the tensor polarizations by the name ``tensor'' memory effects and those associated with the breathing polarization by ``scalar'' memory effects.\footnote{The terminology ``scalar'' and ``tensor'' memory effects also has been used after Du and Nishizawa~\cite{Du:2016hww} by Satishchandran and Wald~\cite{Satishchandran:2019pyc} in the context of general relativity to refer to different classes of ``ordinary'' memory effects (in the sense of~\cite{Bieri:2013ada}); since we work in Brans-Dicke theory, and we compute ``null'' memory effects throughout this paper, we think the sense in which we use this naming should be clear (with this footnote as an attempt to dispel any potential lingering ambiguities).} 
We~\cite{Tahura:2020vsa}, as well as Hou and Zhu~\cite{Hou:2020wbo}, derived the conserved charges associated with the BMS symmetries using the Wald-Zoupas prescription, and we determined that the charges  included contributions coming from the scalar field.

Because~\cite{Tahura:2020vsa} (and~\cite{Hou:2020tnd,Hou:2020wbo}) provided the necessary framework in which to compute and interpret the GW memory effects in BD theory, we now turn to applying the results of~\cite{Tahura:2020vsa} to construct the GW memory waveform from compact binary systems (as one application of this formalism).
We will focus on the memory effects that appear in the tensor polarizations of the GWs, because we can use the BMS flux balance laws to construct nonlinear memory effects based on linearized (or nonlinear) waveforms that do not include the memory effects.
The procedure in Brans-Dicke theory is closely analogous to that used in GR~\cite{Nichols:2017rqr,Nichols:2017rqr,Ashtekar:2019viz,Compere:2019gft,Mitman:2020bjf,Mitman:2020pbt,Khera:2020mcz,Liu:2021zys}. 
The two new memory effects in the scalar polarizations of the GWs are  related to shifts in the scalar field and its time integral.
It was recently shown in~\cite{Seraj:2021qja} that the scalar memory effects are closely related to the large gauge symmetries of 2-form theory that was shown in~\cite{Campiglia:2018see} to be dual to the scalar field theory.
The symplectic flux of the scalar field is linear in the field and in the large gauge transformation; as a result, we cannot use the flux balance laws to construct a nonlinear memory of the scalar waves as one can for the tensor waves via the BMS flux-balance laws (one must instead solve the scalar field equation directly to determine the scalar memory effect).
Since our focus is on the application of the BMS balance laws in BD theory to determine the GW memory effects, we will focus here on computing the tensor memory effects in BD theory, which differ from those of GR due to the emission of scalar radiation. 

In BD theory, tensor GW memory effects are generated by energy and angular momentum fluxes of both tensor and scalar radiation. 
Because solar-system experiments~\cite{Bertotti:2003rm} and pulsar observations~\cite{Freire:2012mg,Archibald:2018oxs} have constrained the amount of scalar radiation in BD theory, we assume that the scalar radiation leads to energy and angular momentum fluxes that are small compared to the leading quadrupole fluxes of tensor GWs in GR. 
Note, however, that the scalar field's fluxes appear at a lower post-Newtonian (PN) than the tensor GW fluxes do (see, e.g.,~\cite{Blanchet:2013haa} for a review of the post-Newtonian, as well as the multipolar post-Minkowskian, expansion).
For a fixed value of the small (dimensionless) inverse coupling parameter in BD theory, there is thus a smallest PN parameter at which our approximation of the smaller scalar-field fluxes holds.
To compute GW memory effects in BD theory at Newtonian order, we will need to include higher-PN-order terms (in the frequency evolution and Kepler's law, for example) than we would need to go to Newtonian order in the calculation in GR.
In addition, we will also truncate our results at a finite, but smallest PN parameter, which is the smallest value for which our approximation holds (unlike in GR, in which we can take the PN parameter to zero).\footnote{Note, of course, that we could also compute the memory effects from a PN parameter of zero up to the small PN parameter at which the fluxes of scalar and tensor radiation have comparable magnitudes, if we assume that the radiated fluxes are dominated by the scalar emission.
This, in fact, is the approximation used in~\cite{Lang:2013fna,Lang:2014osa}, for example.
However, because memory effects are most important when the fluxes are large, this early-time (or small-PN-parameter) regime is not expected to produce a significant GW memory effect, and we do not compute it in this paper.}

We computed our memory effects using the oscillatory waveforms computed in, e.g.,~\cite{Will:1989sk,Lang:2013fna,Lang:2014osa} after verifying that we can relate the waveforms computed in harmonic coordinates in these references to our Bondi-Sachs quantities.
The memory effects that we compute in BD theory, have small terms (proportional to the small BD parameter) that appear at a PN order less than the leading Newtonian order. 
We can relate part of our results to a part of the waveform computed by Lang in~\cite{Lang:2013fna,Lang:2014osa} using the direct integration of the relaxed Einstein equations for the scalar and tensor waveforms up to $1.5$PN and $2$PN orders, respectively. 
Lang found no scalar GW memory effects, but he computed a (hereditary) tensor GW memory effect formally at 1.5PN order that arises from the flux of energy radiated in the scalar waves. 
Upon integrating this 1.5PN term for compact-binary source in our approximation, this term leads to a memory effect that depends logarithmically on the PN parameter (this is analogous to how a formally 2.5PN order term in GR, when integrated for compact binaries, leads to a Newtonian-order effect in the waveform~\cite{Wiseman:1991ss,Favata:2008yd}).
If we compare our result in the Bondi-Sachs framework with Lang's harmonic-coordinate expression, the two terms agree.
The BMS flux balance laws are not as helpful for verifying the absence of scalar memory effects at $1.5$PN order.

Our BMS flux-balance approach allows us to compute the Newtonian-order tensor waveforms---which have not been computed before, as far as we are aware---that should appear if the work of Lang~\cite{Lang:2013fna} were extended to 2.5PN order.
We find that because of the dipole emission, the Newtonian-order GW memory effects sourced by the tensor-GW energy flux has contributions from current quadrupole, mass octopole, and mass hexadecapole moments. 
These higher multipole moments produce GW memory waveforms that have a different dependence on the inclination angle than the tensor GW memory effect in GR at the equivalent PN order. 
The Newtonian GW memory effect generated by the scalar field's energy flux also has a different dependence on inclination angle from that sourced by the tensor GWs. 
The inclination-angle dependence of the GW memory effect has been shown to be something that can be tested with second- and third-generation ground-based GW detectors~\cite{Yang:2018ceq}.

The rest of the paper is organized as follows. 
In Sec.~\ref{sec:harmonic&bondi}, we present a few elements of BD theory in harmonic and Bondi coordinates. 
Section~\ref{sec:expnasionPN&Mulipolar} lists the oscillatory radiative mass and current multipole moments for a quasi-circular, nonspinning compact-binary inspiral (Sec.~\ref{subsec:mass&currentmoments}); it reviews the derivation of Kepler's law, the evolution of the orbital frequency, and the phase of GWs in BD theory at the necessary PN orders in our approximation (Sec. \ref{subsec:kepler'slaw-phase}); and it presents scalar multipole moments generated by an inspiraling quasi-circular, nonspinning compact binary (Sec.~\ref{subsec:scalarmoments}). 
In Sec.~\ref{sec:memorywaveforms}, we compute the nonlinear displacement and spin GW memory waveforms in BD theory from the BMS fluxes. 
We conclude in Sec.~\ref{sec:conclusions}. 
We give additional results in two appendices in which we show the coordinate transformations that relate the Bondi coordinates to harmonic coordinates including relations between the metric functions in two coordinates systems (Appendix~\ref{app1}), and we argue that the ordinary parts of the GW memory effects are subleading compared to null memory effects in BD theory (Appendix~\ref{Sec:Ordinary_Memory}).

Throughout this paper, we use units in which $c=1$, and we also set the asymptotic value of the gravitational constant in BD theory to $1$. We use Greek indices ($\mu,\nu,\dots$) to denote four-dimensional spacetime indices, and uppercase Latin indices $(A, B, C, \dots)$ for indices on the 2-sphere, and lowercase Latin indices ($i,j,k,\dots$) for the spatial indices in quasi-Cartesian harmonic coordinates.

\section{Waveform in harmonic and Bondi coordinates} \label{sec:harmonic&bondi}

In this section, we discuss briefly the Bondi-Sachs framework~\cite{Bondi:1962px,Sachs:1962wk} and the harmonic-gauge waveform in post-Newtonian theory, both of which we will use to compute the GW memory waveform.
Specifically, in Sec.~\ref{subsec:harmonic}, we discuss BD theory in harmonic coordinates and decompose the GW strain into radiative multipole moments. 
In Sec.~\ref{subsec:bondi}, we present BD theory in the Bondi-Sachs framework and again perform a multipole decomposition of the radiative data.
The last part of this section (Sec.~\ref{subsec:Relation}) relates the multipole moments of the shear tensor in Bondi coordinates to the radiative mass and current multipole moments of the harmonic-gauge waveform; it then does the same for the multipole expansion of the scalar waveform in Bondi coordinates and harmonic coordinates.

We require both coordinate systems and frameworks, because the nonlinear and null GW memory effects are straightforward to compute through the BMS balance laws in the Bondi approach, but it is more challenging to relate the Bondi-Sachs framework to a specific solution of a Cauchy initial-value problem.
In the harmonic-gauge PN approach, the scalar and tensor GW waveforms already have been computed generally and for specific compact-binary sources in, e.g.,~\cite{Lang:2013fna,Lang:2014osa}; however, the GW memory effects are of a sufficiently high PN order in PN theory that they have not been fully computed in BD theory.
After relating the harmonic-gauge waveform to the shear in the Bondi-Sachs framework, we can then determine the GW memory waveforms using the balance laws (and thereby avoiding high PN-order calculations).

Throughout this paper, we treat Brans-Dicke theory in the Jordan frame~\cite{Brans:1961sx}, in which the action takes the form
\begin{equation}
S = \int d^{4} x \sqrt{-g}\left[\frac{\lambda}{16 \pi} \mathcal R-\frac{\omega}{16 \pi} g^{\mu \nu} \frac{\left(\partial_{\mu} \lambda\right)\left(\partial_{\nu} \lambda\right)}{\lambda}\right]\,. 
\end{equation}
Here $g_{\mu\nu}$ is the Jordan-frame metric, $\mathcal R$ is the Ricci scalar of $g_{\mu\nu}$, $\lambda$ is a massless scalar field with a nonminimal coupling to gravity, and $\omega$ is a coupling constant called the Brans-Dicke parameter. 
In this section and subsequent ones, we set the gravitational constant at infinity to unity, i.e.,
\begin{equation}
    G_0= \frac{4+2\omega}{3+2\omega} \frac 1{\lambda_0} = 1 \, ,
\end{equation}
where $\lambda_0$ is the constant value that $\lambda$ approaches in the limit of infinite distances from an isolated source.

\subsection{Waveform in harmonic coordinates}\label{subsec:harmonic}

We will denote our quasi-Cartesian harmonic-gauge coordinates by $X^\mu$, and we will use the notation $X^0 = t$ for the time coordinate and $X^i$ (for $i=1$, 2, 3) for the spatial coordinates.
We will denote the Euclidean distance from the origin at fixed $t$ by $R = \sqrt{X^i X^j \delta_{ij}}$.
The tensor GWs in Brans-Dicke theory are described by the transverse-tracelesss (TT) components of the metric perturbation, $\tilde h_{ij}^{\mathrm{TT}}$, and the scalar GWs are encapsulated in the scalar field $\lambda$. 
Both fields can be obtained from the metric at order $1/R$, in an expansion in inverse $R$, from the spatial components of the spacetime metric $g_{ij}$.
The metric is more conveniently written in terms of the metric perturbation $\tilde h_{ij}$ and its trace $\tilde h$ rather than the TT part.
For extracting the GWs, we need only the part of the spacetime metric that is linear in the fields $\lambda$ and $\tilde h_{ij}$ at linear order $1/R$. 
We write the metric in this approximation as in~\cite{Lang:2013fna}, 
\be
g_{ij} = \delta_{ij} + \tilde h_{ij} -\frac{1}{2} \tilde h \delta_{ij} - \lb \frac{\lambda}{\lambda_0} - 1\rb \delta_{ij} \, .
\ee
The scalar GWs are present in the $1/R$ part of $\lambda$, which we expand as
\be\label{eq:lambda_expansion_harmonic}
\lambda=\lambda_0+\frac{\Xi(\tilde u,y^A)}{R} + O \left(R^{-2}\right) \,.
\ee
We have written the scalar field in terms of $\tilde u=t-R$, the retarded time in harmonic coordinates, and the angles $y^A\equiv (\iota,\varphi)$.
The angle $\iota$ is the polar angle and $\varphi$ is the azimuthal angle of a spherical polar coordinate system.\footnote{For compact binary sources, $\iota$ is the inclination angle between the orbital angular momentum of the binary (assumed to be along the $Z$ axis) and $\varphi$ is the azimuthal angle as measured from the $X$ axis.}
We expand the TT projection of $\tilde h_{ij}$ in terms of second-rank electric-parity and magnetic-parity tensor spherical harmonics ($T_{i j}^{(e), l m}$ and $T_{i j}^{(b), l m}$, respectively; see, e.g.,~\cite{Thorne:1980ru}) as~\cite{Blanchet:2013haa}
\be\label{eq:hToUV}
\tilde h_{i j}^{\mathrm{TT}}=\frac{1}{R} \sum_{l, m}\left[U_{l m} (\tilde u) T_{i j}^{(e), l m}+V_{l m} (\tilde u) T_{i j}^{(b), l m}\right] \,.
\ee
The sum runs over integer values of $l$ and $m$ with $l\geq 2$ and $-l\leq m \leq l$.
The coefficients $U_{lm}$ and $V_{lm}$ are two sets of radiative multipole moments which are called the mass and current moments, respectively. 
Because $\tilde h_{ij}^\mathrm{TT}$ is real, the mass and current moments satisfy the following properties under complex conjugation: 
\ba 
\bar{U}_{l m}=(-1)^{m} U_{l,-m}, \quad \bar{V}_{l m}=(-1)^{m} V_{l,-m} \, .
\ea
We use an overline to denote the complex conjugate.

We will also use the complex gravitational waveform $h$ which is composed of the plus and cross polarizations as follows:
\be\label{eq:WaveformPlusCross}
h=h_+ -i h_{\times} \, .
\ee
We use the conventions for the polarization tensors $e^+_{ij}$ and $e^\times_{ij}$ given in~\cite{Favata:2008yd} or~\cite{Kidder:2007rt} to construct the polarizations $h_+ = \tilde h^{ij}_\mathrm{TT} e^+_{ij}$ and $h_\times = \tilde h^{ij}_\mathrm{TT} e^\times_{ij}$.
We expand $h$ as in terms of spin-weighted spherical harmonics ${}_sY_{lm}$ with spin weight $s=-2$:
\be\label{eq:H}
h=\sum_{l, m} h_{l m}(\tilde u) (\prescript{}{-2}{Y}_{l m})\,.
\ee
For a nonspinning planar binary, the modes $h_{lm}$ are related to the mass and current multipole moments by (see, e.g.,~\cite{Blanchet:2013haa})
\be \label{eq:Hlm}
h_{lm}=\begin{cases}
\frac{1}{\sqrt{2}R} U_{lm}& (\text{$l+m$ is even})\,, \\
-\frac{i}{\sqrt{2}R}V_{lm}& (\text{$l+m$ is odd})\,.
\end{cases}
\ee

\subsection{Waveform and metric in Bondi coordinates}\label{subsec:bondi}

We use $(u,r,x^A)$ to denote Bondi coordinates.
The coordinate $u$ is the retarded time, $r$ is an areal radius, and $x^A$ are coordinates on a 2-sphere cross sections of constant $u$ and $r$ (where $A=1$, 2). 
We expand $\lambda$ as a series in $1/r$ as
\begin{align}
\lambda(u,r,x^A) = {} & \lambda_0 + \frac{\lambda_1\left(u,x^A\right)}{r} + O(r^{-2}) \, .
\end{align}
The metric in Bondi gauge satisfies the conditions $g_{rr} = 0$, $g_{rA}=0$, and the determinant of the metric on the 2-sphere cross sections scaled by $r^{-4}$ is independent of $r$ (and $u$).
We imposed a set of aysmptotic boundary conditions on the nonzero components of the metric in Bondi gauge in~\cite{Tahura:2020vsa} and postulated a Taylor series expansion of the scalar field and metric on the 2-sphere cross sections in $1/r$. 
This allowed us to solve the field equations of Brans-Dicke theory to obtain the following solution for the line element~\cite{Tahura:2020vsa}:
\ba\label{eq:BondiMetric}
ds^2 &&=-\left[1+\frac{\dot\lambda_1}{\lambda_0}\right.
\left.-\frac{1}{r}\lb 2\mathcal M+\frac{\lambda_1}{\lambda_0}+\frac{3\lambda_1}{2\lambda_0^2}\dot \lambda_1\rb\right]du^2 \nonumber \\
&&-2\left(1-\frac{\lambda_1}{\lambda_0 r} \right)du dr+r^2\left(q_{AB}+\frac{1}{r}c_{AB}\right)dx^A dx^B\nonumber\\
&&+\left\{ \eth_F c^{AF}-\frac{\eth^A \lambda_1}{\lambda_0}+\frac{1}{r}\left[-4 L_{A}+\frac{1}{3} c_{A B}\eth_C c^{B C}
\right.\right.\nonumber\\&&-\left.\left.\frac{1}{3 \lambda_0}\left(2 \lambda_1\eth^B c_{AB}+c_{A B}\eth^{B} \lambda_1-\frac{1}{\lambda_0}\eth_A \lambda_1^2\right)\right]\right\} du dx^A
\nonumber\\&&+\ldots\,.
\ea
The ellipsis at the end of the equation indicates higher order terms in powers of $1/r$ that we are neglecting (the terms are of order $1/r^2$ except for the term proportional to $dx^A dx^B$, which is of order unity, because of the $r^2$ term multiplying the expression). 
In Eq.~\eqref{eq:BondiMetric}, we have introduced $\mathcal M$ and $L_A$ which are (related to) functions of integration in Brans-Dicke theory that are the analogues of the Bondi mass aspect and angular momentum aspect in GR~\cite{Tahura:2020vsa}. 
The two-dimensional metric $q_{AB}$ is the unit-sphere metric and $\eth^A$ is the covariant derivative compatible with $q_{AB}$. 
We will raise and lower 2-sphere indices (such as $A$ and $B$) with the metrics $q^{AB}$ and $q_{AB}$, respectively.
The overhead dot means a partial derivative with respect to the retarded time $u$.
The symmetric trace-free tensor $c_{AB}$ is called the shear tensor, and is related to the GW strain. 
The time-derivative of $c_{AB}$ is a symmetric trace-free tensor known as the news tensor:
\be \label{eq:NABfromCAB}
N_{AB}=\partial_u c_{AB}\,.
\ee
It is not constrained by the asymptotic field equations in Brans-Dicke theory, and it contains information about the tensor GWs.
In GR, if the news tensor vanishes it means the corresponding region of spacetime contains no GWs~\cite{Geroch:1977jn}.

We will also expand $c_{AB}$ in spherical harmonics as 
\ba\label{eq:cToUV}
c_{A B}=\sum_{l,m}\left(c_{(e), l m} T_{A B}^{(e), l m}+c_{(b), l m} T_{A B}^{(b), l m}\right)\,.
\ea
The tensor spherical harmonics can be defined from the scalar spherical harmonics
\begin{subequations}
\begin{align}
    \label{eq:TensorSphElectric}
    T_{A B}^{(e), l m} = {} & \frac{1}{2} \sqrt{\frac{2(l-2)!}{(l+2)!}}\left(2 \eth_{A} \eth_{B}-q_{A B} \ETH^{2}\right) Y_{l m} \, , \\
    \label{eq:TensorSphMagnetic}
    T_{A B}^{(b), l m} = {} & \sqrt{\frac{2(l-2) !}{(l+2) !}} \epsilon_{C(A} \eth_{B)} \eth^{C} Y_{l m} \, ,
\end{align}
\end{subequations}
or instead in terms of spin-weighted spherical harmonics and a complex null dyad on the unit 2-sphere of $m^A$ and its complex conjugate $\bar m^A$ (see, e.g.,~\cite{Thorne:1980ru}):
\begin{subequations}
\label{eq:TABto2Ylm}
\begin{align}
    \label{eq:TensorSphElectricDyads}
    T_{A B}^{(e), l m} = {} & \frac{1}{\sqrt{2}}\left(_{-2}Y_{l m} m_{A} m_{B}+{ }_{2} Y_{l m} \bar{m}_{A} \bar{m}_{B}\right) \, , \\
    \label{eq:TensorSphMagneticDyads}
    T_{A B}^{(b), l m} = & -\frac{i}{\sqrt{2}}\left(_{-2}Y_{l m} m_{A} m_{B}-{ }_{2} Y_{l m} \bar{m}_{A} \bar{m}_{B}\right) \, .
\end{align}
\end{subequations}
The dyad is normalized such that $m^A \bar m_A = 1$.

\subsection{Relation between Bondi- and harmonic-gauge quantities}
\label{subsec:Relation}

We construct a coordinate transformation between harmonic and Bondi gauges in Appendix~\ref{app1} that brings the harmonic metric at order $1/R$ to a Bondi-gauge metric at an equivalent order. 
The procedure used is similar to that recently outlined in~\cite{Blanchet:2020ngx}, but it is adapted to Brans-Dicke theory (rather than GR) and it is accurate only to the first nontrivial order in $1/R$.
This coordinate transformation leads to a simple relationships between $c_{AB}$ and $\tilde h_{ij}^{\mathrm{TT}}$, first, and $\lambda_1$ and $\Xi$, second:
\begin{subequations}
\label{eq:BondiHarmonicRadiativeData}
\begin{align}\label{eq:ctohbar}
c_{AB}(u,x^A) = {} & R\,\tilde h_{ij}^{\mathrm{TT}}(t-R,y^A)\partial_A n^i \partial_B n^j\,,\\
\lambda_1(u,x^A) = {} & \Xi(t-R,y^A)\,.
\end{align}
\end{subequations}
The spatial vector $n^i$ is the unit vector pointing radially outward at fixed $t$ in harmonic coordinates (i.e., in the direction of propagation for outgoing GWs). 
The full transformation between the spherical polar coordinates $R$ and $y^A = (\iota,\varphi)$ constructed from the quasi-Cartesian harmonic coordinates and Bondi coordinates is given in App.~\ref{app1}; we list here the relevant leading-order parts of the transformation needed to relate $u$ to $t-R$, $r$ to $R$, and $x^A$ to $y^A$ in Eq.~\eqref{eq:BondiHarmonicRadiativeData}: 
\begin{subequations}
\begin{align}
\label{eq:uBfromuH}
u = {} & t - R - \frac{2M}{\lambda_0} \log(R) + O(R^{-1}) \, , \\
r = {} & R + O(R^0) \, , \qquad x^A = y^A + O(R^{-2}) \, .
\end{align}
\end{subequations}

The second-rank tensor spherical harmonics on the unit 2-sphere in spherical and Cartesian coordinates are related by the following transformation:
\begin{subequations}
\ba
&T_{AB}^{(e), l m}=T_{ij}^{(e), l m}\partial_A n^i \partial_B n^j\,,\\
&T_{AB}^{(b), l m}=T_{ij}^{(b), l m}\partial_A n^i \partial_B n^j\,.
\ea
\end{subequations}
Combining the expressions~\eqref{eq:hToUV}, \eqref{eq:cToUV}, and \eqref{eq:ctohbar}, we find that the multipole moments of the strain and shear are related by
\be \label{eq:clmtoUV}
c_{(e), l m}=U_{l m}, \quad c_{(b), l m}=V_{l m}\, , 
\ee
as was given in~\cite{Nichols:2017rqr} (though there the relationship was derived through a different argument involving the Riemann tensor in linearized gravity).
The relation~\eqref{eq:clmtoUV} allows us to express the multipole moments of the shear tensor in terms of multipole moments of the harmonic-gauge TT strain tensor, once the difference between the retarded times in harmonic and Bondi coordinates in Eq.~\eqref{eq:uBfromuH} is taken into account.

\section{Post-Newtonian radiative multipole moments}\label{sec:expnasionPN&Mulipolar}

In this section, we compute expressions for the radiative multipole moments $U_{lm}$ and $V_{lm}$, as well as the scalar multipole moments which we will define herein. 
We obtain the moments for nonspinning, quasicircular binaries.
We denote the total mass by $M=m_1+m_2$ (where $m_1$ and $m_2$ are the individual masses), the symmetric mass-ratio by $\eta=m_1m_2/M^2$, and orbital separation by $a$. 
We also introduce the parameters $\xi = 1/(2+\omega)$ and $x = (\pi M f)^{2/3}$, where $f$ is the GW frequency. 
We will work in the approximation in which $\xi\ll x$, which corresponds to assuming that the BD modifications to the waveforms and the dynamics are small corrections to the corresponding quantities in GR. 
Given that the Shapiro-delay measurement in the solar system bounds the BD parameter to be $\omega > 4 \times 10^4$~\cite{Bertotti:2003rm} (a similar bound has been derived from the pulsar triple system PSR J0337+1715~\cite{Archibald:2018oxs}), this implies that our approximation is valid when $x \gg 2.5 \times 10^{-5}$.
In this paper, we will compute GW memory waveforms through Newtonian order, and we keep BD terms that are linear in $\xi$.
We then retain only the terms in the radiative multipole moments at the appropriate powers of $x$ and $\xi$ to obtain Newtonian-order-accurate memory waveforms.
Because we always work to linear order in $\xi$, we do not include error terms of $O(\xi^2)$ in our expressions (they should be considered to be implied).
We do, however, include such error terms in the PN parameter $x$, because the power of $x$ that constitutes a ``Newtonian-order quantity'' is not the same for the various quantities that we consider in the next two sections.

We will need radiative mass and current multipole moments at a higher PN order than those required for computing 0PN memory effects in GR. 
Specifically, in GR, one only requires the mass quadrupole moment at Newtonian order---i.e., $O(x)$---to obtain the Newtonian-order memory effects. 
In BD theory, we will need several additional terms.
First, we must compute the BD correction linear in $\xi$ to the mass quadrupole moment. 
Second, we will need the 1PN or $O (x^2)$ GR terms of the mass quadrupole moment, because they multiply $-1$PN terms present in the GW phase and in the frequency evolution to contribute to Newtonian-order memory effects. 
Third, we require the current quadrupole and mass octupole moments which begin at 0.5PN order or $O (x^{3/2} )$.
Fourth, we must include the current octopole and mass hexadecapole moments at 1PN or $O(x^2)$. 
We do not need to compute the linear in $\xi$ BD corrections to the $O (x^{3/2} )$ and $O (x^2 )$ radiative multipole moments, however, because the Newtonian-order GW memory effects produced by them would be quadratic in $\xi$. 
For scalar multipole moments, we require the dipole with relative 1PN corrections [up to $O(x^{3/2})$], the quadrupole at $O(x)$ and the octopole, which begins at $ O(x^{3/2})$.
The scalar field moments are all linear in $\xi$.

\subsection{Radiative mass and current multipole moments}\label{subsec:mass&currentmoments}

We first give the expression of lowest-order radiative mass multipole moment $U_{22}$. 
We decompose $U_{22}$ into two parts
\be
U_{22} = U_{22}^0 + U_{22}^\mathrm{1,GR}\, ,
\ee
where $U_{22}^0$ is the Newtonian-order part of $U_{22}$, which includes the BD corrections to linear order in $\xi$, and $ U_{22}^\mathrm{1,GR}$ consists of the 1PN terms in $U_{22}$ in GR. 
We can obtain the moment $U_{22}$ from the expression for $\tilde h^{\mathrm{TT}}_{ij}$ in BD theory in Eqs.~(7.1) and~(7.2a) of Ref.~\cite{Lang:2013fna} by
contracting $\tilde h^{\mathrm{TT}}_{ij}$ with $\bar T_{(e),22}^{ij}$ and integrating over the 2-sphere.
We find that to linear order in $\xi$,
\be
U_{22}^0 =-8 \sqrt{\frac{2\pi}{5}}  \eta M x \,e^{-i \phi} \lb 1-\frac{\xi}{2} - \frac 23 \mathscr G \rb\, , 
\ee
where 
\begin{equation}
    \mathscr G = \xi(s_1+s_2-2s_1s_2) \, .
\end{equation}
The variables $s_1$ and $s_2$ denote the sensitivities (see, e.g.,~\cite{Will:1989sk}) of the binary components. 
The quantity $\mathscr G$ is related to the modified gravitational constant in BD theory  by
\begin{equation} \label{eq:CalG}
\mathcal{G} = 1 - \mathscr G \, .
\end{equation} 

The GW phase is denoted $\phi(x)$, and it differs from the phase of the $l=2$, $m=\pm 2$ modes of the waveform in GR; we give the expression for the phase in Eq.~\eqref{eq:phase} below.
We use the 1PN GR terms from the review article~\cite{Blanchet:2013haa}.
Putting these two results together, we have the following expression for $U_{22}$:
\begin{subequations}
\label{eq:radiativeUVmoments}
\begin{align}
\label{eq:U22}
U_{22} = {} & -8 \sqrt{\frac{2\pi}{5}}  \eta M x \,e^{-i \phi} \nonumber \\ & \times \left[\left(1 - \frac{\xi}{2} -\frac 23 \mathscr G \right)  +\left(\frac{55 \eta }{42}-\frac{107}{42}\right) x \right] + O(x^{5/2})\,. 
\end{align}
The term proportional to $x$ in the square bracket is the 1PN GR term taken from~\cite{Blanchet:2013haa}.

As we will show in Section~\ref{sec:memorywaveforms},
to compute the GW memory waveform at Newtonian order, we need the radiative current quadrupole moment and several radiative mass octopole and hexadecapole moments.
To work to linear order in $\xi$, we can use the GR amplitudes of the moments (though we use the phase with the BD corrections).
This allows us to take the amplitudes from the expressions given, e.g., in the review~\cite{Blanchet:2013haa}:
\begin{align}
\label{eq:V21}V_{21} = {} & \frac{8}{3} \sqrt{\frac{2 \pi}{5}} \eta \delta m \,x^{3/ 2} e^{-i \phi/2}  + O(x^{5/2}) \,,\\
U_{33} = {} & 6 i\sqrt{\frac{3 \pi}{7}} \eta \delta m\, x^{3/ 2} e^{-3i \phi/2} + O(x^{5/2}) \,,\\
V_{32} = {} & i \frac{8}{3} \sqrt{\frac\pi{14}}M \eta (1-3\eta) x^2 e^{-i \phi} + O(x^{5/2}) \,,\\
U_{31} = & -\frac{2i}{3}\sqrt{\frac{\pi}{35}} \eta \delta m \,x^{3/ 2} e^{-i \phi/2} + O(x^{5/2}) \,,\\
\label{eq:U42}
U_{42} = & -\frac{8}{63} \sqrt{2 \pi}M \eta (1-3\eta) x^2 e^{-i \phi} + O(x^{5/2}) \,.
\end{align}
\end{subequations}
We use the notation $\delta m=(m_1 -m_2)$.
We give an expression for the phase $\phi(x)$ in the next subsection.

\subsection{Kepler's law, frequency evolution, and GW phase}\label{subsec:kepler'slaw-phase}

Before computing the phase, we first give an expression for Kepler's law, which we will need to compute the scalar multipole moments and the frequency evolution as well as the phase. 
To obtain a Newtonian-order accurate GW memory waveform, we need to have an expression for Kepler's law at 1PN order.
This higher order is needed, because when evaluating the integrals involved in computing the GW memory effect, there are $-1$PN terms arising from dipole radiation in the energy flux, which multiply 1PN terms in the GW frequency's evolution and give rise to Newtonian-order terms in the waveform.
The two-body equations of motion of nonspinning compact objects in Brans-Dicke theory has been computed in Ref.~\cite{Mirshekari:2013vb}.
For circular orbits, the relative acceleration is proportional to the orbital frequency squared, $\Omega^2$, and the relative separation to 1PN order.
Working to linear order in $\xi$, the results of Eqs.~(1.4) and (1.5a) of~\cite{Mirshekari:2013vb} show that Kepler's law in BD theory (in this approximation) is
\be \label{eq:Kepler'sLaw}
\Omega^2=\frac{M}{a^3}\left[1-\mathscr G -\frac{M}{a} \left(1-2\mathscr G \right)\left(3-\eta\right) - \frac Ma \mathcal G\bar\gamma \right] \, .
\ee 
We have introduced the parameter
\begin{equation}
\bar \gamma=-\mathcal G^{-1} \xi\left(1-2 s_{1}\right)\left(1-2 s_{2}\right) \, ,
\end{equation} 
in the equation above and $\mathcal G$ is defined in Eq.~\eqref{eq:CalG}.

Let us now compute the evolution of the GW frequency. 
We again need a 1PN-order-accurate expression, which, in full generality, contains many terms. 
Because we work to linear order in $\xi$, the only 1PN terms that we need to obtain a Newtonian-order expression for the GW memory waveforms are the GR terms in $\dot f$ (i.e., the 1PN terms without $\xi$).\footnote{The $-1PN$ term that multiplies the 1PN term in this calculation is linear in $\xi$, which implies that the BD modification to $\dot f$ at 1PN enters at higher order in $\xi$ in the GW memory waveforms.}
We can then write the expression for $\dot f = df/dt$ in the following form:
\begin{equation}
    \dot f = \dot f_0 + \dot f_\mathrm{1,\mathrm{GR}} \, ,
\end{equation}
where $\dot f_0$ is the BD expression to Newtonian order and linear order in $\xi$, and $\dot f_\mathrm{1,GR}$ is $\xi=0$ (or GR) limit of the 1PN terms.
We first compute $\dot f_0$ using results from~\cite{Mirshekari:2013vb}, and then we add to it the terms $\dot f_\mathrm{1,GR}$ taken from~\cite{Arun:2004hn}.
To compute $\dot f_0$, we first use the binding energy of a binary in BD theory from Eq.~(6.14) of~\cite{Mirshekari:2013vb}, which is valid to 1PN order:
\begin{align}\label{eq:Binding_Energy}
E_b = {} & \frac{1}{2} \mu v^{2}-\mu \frac{\mathcal G M}{a}+\frac{3}{8} \mu(1-3 \eta) v^4 \nonumber \\
& +\frac{1}{2} \mu \frac{\mathcal G M}{a}(3+2 \bar{\gamma}+\eta) v^2 
+\frac{1}{2} \mu(1-2\mathscr G) \left(\frac{M}{a}\right)^{2} \, .
\end{align}
We used $\mu = \eta M$ to denote the reduced mass.
We will next express the binding energy in terms of the PN parameter $x=(\pi M f)^{2/3}$.
To do this it is useful to have the expressions for $M/a$ and $v^2$ written in terms of $x$:
\begin{subequations}
\label{eq:Ma_v_x}
\begin{align}
    \frac M a = {} & x \left[ 1 + \frac 13 \mathscr G + \left(1-\frac 13 \mathscr G \right)\left(1 - \frac 13 \eta\right) x + \frac 13 \mathcal G \bar \gamma x\right] \nonumber \\
    & + O(x^{3}) \, , \\
    v = {} & \sqrt x \left[ 1 - \frac 13 \mathscr G - \left(1-\mathscr G \right)\left(1 - \frac 13 \eta\right) x - \frac 13 \mathcal G \bar \gamma x\right] \nonumber \\
    & + O(x^{5/2}) \, .
\end{align}
\end{subequations}
We can then substitute Eq.~\eqref{eq:Ma_v_x} into Eq.~\eqref{eq:Binding_Energy} to obtain
\begin{align}
    E_b = & -\frac 12 \mu x\left[1 - \frac 23 \mathscr G - \frac 1{12}\left(1-\frac 43\mathscr G\right)(9+\eta)x - \frac 23 \mathcal G \bar \gamma x \right] \nonumber \\
    & + O(x^3) \, .
\end{align}

The rate of change of energy radiated in GWs in BD theory through Newtonian order is given by a $-1$PN term plus a Newtonian term.
If we define 
\begin{equation}
    S  =s_1 - s_2 \, 
\end{equation}
and we make use of expressions (6.16) and (6.19) given in~\cite{Mirshekari:2013vb}, then linearizing their expression in $\xi$, we have
\begin{align}\label{eq:Energy_Flux}
    \dot E_\mathrm{GW} = {} & \frac{32}5 \eta^2 x^5 \bigg[\frac{5 \xi S^2}{48 x} +  1-\frac 73 \mathscr G + \frac{5}{12} \mathcal G\bar\gamma \nonumber \\
    & - \frac{5}{72}\xi S^2(3+2\eta)\bigg] + O(x^{11/2}) \, .
\end{align}

Imposing energy balance $\dot E_b=-\dot E_{\mathrm{GW}}$ (the change in the binding energy is equal to the energy radiated by the GWs) and using the chain rule to write $\dot f = (df/dE_b) \dot E_b$, we can write the Newtonian-order frequency derivative $\dot f_0$ as a function of the PN parameter $x$ as 
\begin{align}
   \dot f_0 = {} & \frac{96\eta}{5\pi M^2}x^{11/2} \left[ 1 + \xi \left( \frac{5S^2}{48 x} + \mathscr F \right)\right] \,,
\end{align}
where we defined
\begin{align}
\mathscr F = & -\frac{5}{12}-\frac{5}{6}(s_1+s_2)+\frac{5}{144}(51+7\eta)s_1 s_2\nonumber\\
& -\frac{5}{288}(3+7\eta)(s_1^2+s_2^2)\,.
\end{align}
Finally, including the GR frequency evolution at 1PN~\cite{Arun:2004hn} to $\dot f_0$, we find 
\begin{align}\label{eq:fdot}
\dot f = {} & \frac{96\eta x^{11/2}}{5\pi M^2} \left[ 1 + \xi \left( \frac{5S^2}{48 x} + \mathscr F \right)-\lb \frac{743}{336}+\frac{11}{4}\eta\rb x \right] \nonumber \\ 
& + O(x^{7}) \,.
\end{align}

We previously introduced a waveform phase variable $\phi(x)$, which we will now compute explicitly.
For computing the GW memory waveforms, we will again need an expression for the GW phase through 1PN order; however, because we are working to linear order in $\xi$, we will only need the terms without $\xi$ at 1PN order in the phase (analogously to our calculation of $\dot f$). 
The GW phase is typically obtained by integrating the GW frequency with respect to time from some appropriate starting time.
For calculations of the GW memory waveform, it is more useful to write the phase as a function of $x$.
By using the chain rule, we can then write the time integral of the frequency in terms of an integral with respect to the PN parameter $x$ as follows:
\begin{equation}
    \phi(x) = 2\pi \int_{x_i}^x \frac{f}{\dot f} \frac{df}{dx'} dx' \, ,
\end{equation}
where the frequency $f$ and the derivatives $\dot f$ and $df/dx$ are functions of $x$.
We have also introduced an initial PN parameter $x_i$ that should be greater than $\xi$, so that our approximation of $\xi \ll x$ holds.
From $\dot f$ in Eq.~\eqref{eq:fdot}, we find that the GW phase is given by 
 \begin{align}
\label{eq:phase}
\phi(x) - \phi_\mathrm{c} &{} = -\frac{1}{16 \eta  x^{5/2}}\left[1-\xi\left( 
\mathscr F +\frac{25 S^2}{336 x}\right) \right.\nonumber \\
& \left. + \frac 53\lb  x - \frac 18 \xi S^2 \rb\lb\frac{743}{336}+\frac{11}{4} \eta \rb \right] + O(x^{-1})\,.
\end{align}
We defined a constant $\phi_\mathrm{c}$, the phase at coalescence, which is chosen such that the phase at $x_i$ vanishes.
The terms in the second line of Eq.~\eqref{eq:phase} come from the product of a $-1$PN
term multiplying a 1PN term, which produces a Newtonian-order effect on the phase (specifically, it is the scalar dipole radiation in BD theory that gives rise to the $-1$PN-order effects). 
We discuss the scalar radiation in more detail in the next subsection.

\subsection{Scalar Multipole Moments}\label{subsec:scalarmoments}

We expand $\lambda_1$ in terms of the scalar spherical harmonics
\be\label{eq:Lambda_Expansion}
\lambda_1=\sum_{l,m} \lambda_{1(lm)}Y^{lm}\, ,
\ee
and the corresponding coefficients in the expansion are the scalar multipole moments $\lambda_{1(l,m)}$.
Specifically, we compute expressions for the scalar moments $\lambda_{1(11)}$, $\lambda_{1(22)}$ and $\lambda_{1(31)}$ in terms of the PN parameter $x$, which we will then use to derive the tensor GW memory effect sourced by the fluxes of the scalar field.
These three moments are those needed to compute the GW memory effects at Newtonian order.
The relevant part of the scalar field $\lambda_1$ had been computed previously in~\cite{Will:1989sk,Lang:2014osa}, and we give the expression to 1PN order above the leading dipole radiation and to linear order in $\xi$:
\begin{align}\label{eq:scalar_field}
\lambda_1 = {} & \eta M \lambda_0 \xi \left\{\left[-2 S+\frac{M}{2 a}\left(3 \Gamma  \frac{\delta m}{M} +10 S  \eta \right) \right]v_i n^i \right.\nonumber\\& \left.
+ \Gamma \left[ (v_i n^i)^2 - \frac{ M}{a} (n_{12}^i n_i)^2 \right] - (\Gamma \frac{\delta m}{M} + 2 \eta S)\right.\nonumber\\ 
&\left. \times\left[(v_i n^i)^3 - \frac{7}{2} \frac{M}{a}(v_i n^i) (n^i_{12} n_i)^2  \right]\right\}\, .
\end{align}
Above we introduced the quantities
\begin{equation}
    \Gamma=1-2(m_1 s_2 + m_2 s_1)/M \, ,
\end{equation}
the unit vector $n^i$ pointing radially outward in the direction of the GW's propagation, the unit separation vector $n_{12}^i$ between the binary's components, and the relative velocity vector $v^i = v_1^i - v_2^i$ of the binary's masses. 
In terms of $\iota$ and $\varphi$ (the polar and the azimuthal angles, respectively, in the center-of-mass frame of the binary) and the GW phase $\phi$, the two unit vectors and the relative velocity vector take the form
\begin{subequations}
\label{eq:various_defs}
\begin{align}
n^{i} = {} & (\sin \iota \cos \varphi, \sin \iota \sin\varphi,\cos\iota)\,, \\
n_{12}^i = {} &  \{ \cos[\phi(u)/2],\sin[\phi(u)/2],0\}\,, \\
\label{eq:v}
v^i = {} &  \{-v\sin[\phi(u)/2],v\cos[\phi(u)/2],0\}\, .
\end{align}
\end{subequations}
For the magnitude of the velocity, $v$, we need only the GR limit of the expression (zeroth-order in $\xi$) at 1PN order in $x$ in Eq.~\eqref{eq:Ma_v_x}.

We wrote the phase as $\phi(u)$ as a shorthand for $\phi[x(u)]$ in Eq.~\eqref{eq:phase}, so as to emphasize the retarded-time dependence of the phase. 
The multipole moments can be extracted through the integral
\begin{equation}\label{eq:lambdaExpansion}
\lambda_{1(lm)} = \int d^2\Omega \, \lambda_1 \bar Y_{lm}(\iota,\varphi)\, .
\end{equation}
Using Eqs.~\eqref{eq:Lambda_Expansion}--\eqref{eq:various_defs} and the GR limit of Eq.~\eqref{eq:Ma_v_x},
we find that the harmonic components of $\lambda_1$ can be written as
\begin{subequations}
\label{eq:scalarmoments}
\begin{align}
\label{eq:lambda111}
\lambda_{1(11)} = & - 2 i \sqrt{\frac{2 \pi }{3}} \lambda_0  \xi  \eta M \sqrt{x} e^{-i \phi/2}  \nonumber\\ 
& \times \left\{S -\frac{x}{15} \left[12\Gamma  \frac{\delta m}{M} +S \left(15 + 34 \eta \right) \right] \right\} + O(x^2) , \\
\lambda_{1(22)} = & -2 \sqrt{\frac{2 \pi }{15}} \lambda_0 \xi \Gamma \eta  M x e^{-i \phi} + O(x^2) \,, \\
\lambda_{1(31)} = & - \frac{i}{10} \sqrt{\frac{\pi }{21}} \lambda_0 \xi \left(\Gamma  \frac{\delta m}{M} + 2 \eta  S\right) \eta  M x^{3/2} e^{-i \phi/2} \nonumber \\ 
& + O(x^2) \,.
\label{eq:lambda22}
\end{align}
\end{subequations}
Equations~\eqref{eq:radiativeUVmoments} and~\eqref{eq:scalarmoments} are all the sets of radiative moments that we will need to compute the GW memory effects in the next section.

\section{Memory Effects}\label{sec:memorywaveforms}

In this section, we compute the displacement and spin GW memory effects produced by a quasicircular compact-binary inspiral. 
The displacement and spin memory effects are both constructed from the shear tensor $c_{AB}$, and they have sky patterns with opposite parities.
It is then useful to first decompose the shear tensor into electric- and magnetic-parity parts as follows:
\be\label{eq:cToThetaPsi}
c_{AB}=\frac{1}{2}(2\eth_A \eth_B-q_{AB}\ETH^2)\Theta+\epsilon_{C(A} \eth_{B)} \eth^{C} \Psi\,,
\ee
where $\ETH^2\equiv\eth^A \eth_A$ is the Laplacian operator on the unit 2-sphere.
We compute the displacement and spin GW memory effects using the BMS flux-charge balance laws that were computed in Brans-Dicke theory in~\cite{Tahura:2020vsa}.
We focus on the nonlinear GW memory effects and the null memory associated with the stress-energy tensor of the scalar waves.
We can compute these effects using low-PN-order oscillatory waveforms and the BMS balance laws, whereas if we were to try to compute them directly through the relaxed Einstein equations in harmonic gauge, we would need to compute the gravitational waveform at a higher PN order in BD theory than has been completed thus far.
We also argue that the ordinary parts of the GW memory effects will be of a higher PN order than the nonlinear and null parts in Appendix~\ref{Sec:Ordinary_Memory}.

\subsection{Spherical harmonics and angular integrals}

We will compute multipole moments of the GW memory effects, starting from the oscillatory tensor and scalar waves expanded in terms of the multipole moments in Eqs.~\eqref{eq:radiativeUVmoments} and~\eqref{eq:scalarmoments}, respectively.
Evaluating these multipole moments involves computing angular integrals involving products of three spherical harmonics of different types (scalar, vector, and tensor).
We instead follow the strategy in, e.g.,~\cite{Nichols:2017rqr,Nichols:2018qac}, in which the vector and tensor harmonics are recast in terms of spin-weighted spherical harmonics.
The angular integrals then involve products of three spin-weighted spherical harmonics (we use the conventions for the spherical harmonics in~\cite{Nichols:2017rqr}). 
We also use the notation for the integral of three spin-weighted spherical harmonics in~\cite{Nichols:2018qac}
\begin{align}
    \label{eq:sYlmToBl}
    & \mathcal B_l(s',l',m';s'',l'',m'') \equiv \nonumber\\
    & \int d^2\Omega \, (_{s'}Y_{l'm'})(_{s''}Y_{l''m''})(_{s'+s''}\bar Y_{l(m'+m'')}) \, ,
\end{align}
which can be written in terms of Clebsch-Gordan coefficients (denoted by $\langle l',m';l'',m''|l,m'+m''\rangle$) as was shown, e.g., in~\cite{Beyer:2013loa} (though using the conventions of~\cite{Nichols:2017rqr}):
\begin{align}\label{eq:Bl}
    \mathcal B_l(s',l',m';s'',l'',m'') = (-1)^{l+l'+l''} \sqrt{\frac{(2l'+1)(2l''+1)}{4\pi(2l+1)}} \nonumber\\
    \times \langle l',s';l'',s''|l,s'+s''\rangle \langle l',m';l'',m''|l,m'+m''\rangle\,.
\end{align}

The multipolar expansion of the nonlinear memory effects in terms of the radiative moments $U_{lm}$ and $V_{lm}$ has the same form as in GR, which is given in~\cite{Nichols:2017rqr}.
However, we will need to perform a new multipolar expansion of the null memory effects from the stress-energy tensor of the scalar field.
For this expansion, we will need the vector spherical harmonics
\begin{subequations}
\ba
\label{eq:VectorSphElectric}
T_{A}^{(e), l m} &=&\frac{1}{\sqrt{l(l+1)}} \eth_{A} Y_{l m}\,,\\
\label{eq:VectorSphMagnetic}
T_{A}^{(b), l m} &=&\frac{1}{\sqrt{l(l+1)}} \epsilon_{A B} \eth^{B} Y_{l m}\, .
\ea
\end{subequations}
In terms of the spin-weighted spherical harmonics and a complex dyad $m^{A}$ on the unit 2-sphere, we can write the vector spherical harmonics as
\begin{subequations}
\ba
\label{eq:VectorSphElectricDyads}
T_{A}^{(e), l m} &=&\frac{1}{\sqrt{2}}\left(_{-1}Y_{l m} m_{A}-{ }_{1} Y_{l m} \bar{m}_{A}\right)\,,\\
\label{eq:VectorSphMagneticDyads}
T_{A}^{(b), l m}&=&\frac{i}{\sqrt{2}}\left(_{-1}Y_{l m} m_{A}+{ }_{1} Y_{l m} \bar{m}_{A}\right)\,,
\ea
\end{subequations}
where $m^Am_A=\bar m^A \bar m_A=0$ and $m^A \bar m_A=1$.

\subsection{Displacement memory effects} \label{subsec:displacementmemory}

Supermomentum conservation requires that the change in the ``potential'' $\Theta$ that is associated with the electric part of the shear tensor, $\Delta\Theta$, must have its change between two retarded times and satisfy the following relationship~\cite{Tahura:2020vsa}:
\begin{widetext}
\begin{align}
\label{eq:DispMemoryFlux}
    \int d^2\Omega \, \alpha \,  \ETH^2 (\ETH^2+2) \Delta \Theta  = \int du \, d^2\Omega \, \alpha \left[ N_{AB} N^{AB} 
     +\frac{6+4\omega}{(\lambda_0)^2} (\partial_u \lambda_1)^2\right]+8\int \, d^2\Omega\,\alpha \left( \Delta \mathcal M - \frac 1{4\lambda_0} \ETH^2 \Delta \lambda_1 \right) \, .
\end{align}
\end{widetext}
The supermomentum is the charge conjugate to a BMS supertranslation symmetry and $\alpha(x^A)$ is the function that parametrizes the supertranslation symmetry. 
The first two terms inside the square brackets on the right-hand side of Eq.~\eqref{eq:DispMemoryFlux} produce the null memory (i.e., the memory sourced by massless fields) with the first term being the nonlinear (Christodoulou) memory.
Both $\Delta\mathcal M$ and $\ETH^2 \Delta\lambda_1$ generate ordinary memory~\cite{Tahura:2020vsa}, but we argue in App.~\ref{Sec:Ordinary_Memory} that the ordinary memory is a higher-PN-order effect. 
We will then focus on just the null memory, and we will derive separately the contributions from the energy flux of tensor and scalar waves, respectively. 
We denote the nonlinear (tensor) part by $\Delta \Theta^\mathrm{T}$ and the null part from the scalar field by $\Delta \Theta^\mathrm{S}$.
The full memory effect is then the sum of the two components:
\begin{equation}
    \Delta \Theta = \Delta \Theta^\mathrm{T} + \Delta \Theta^\mathrm{S} \, .
\end{equation}

While $\Delta\Theta$ is the quantity most straightforwardly constrained by supermomentum conservation, it is the change in the strain $\Delta h$ that is perhaps the more typical gravitational-wave observable.
It is thus useful to relate the potential $\Delta\Theta$ to the strain $\Delta h$.
To do this, we will first introduce the following notation for just the electric part of the change in the shear:
\be\label{eq:cToTheta}
\Delta c_{AB,(e)} = \frac{1}{2}(2\eth_A \eth_B-q_{AB}\ETH^2) \Delta\Theta \,.
\ee
We expand $\Delta \Theta$ in scalar spherical harmonics
\be \label{eq:ThetaExpansion}
\Delta \Theta=\sum_{l,m} \Delta \Theta_{lm} Y^{lm}(\iota, \varphi)\,,
\ee
where $l\geq 2$ (and $-l\leq m \leq l$); the $l\leq 1$ harmonics are in the kernel of the operator $2\eth_A \eth_B-q_{AB}\ETH^2$. 
By substituting~\eqref{eq:ThetaExpansion} into Eq.~\eqref{eq:cToTheta} and using the definition of $T_{A B}^{(e), l m}$ in Eq.~\eqref{eq:TensorSphElectric},
we can relate $\Delta \Theta_{lm}$ to $\Delta c_{(e),lm}$ via Eq.~\eqref{eq:cToUV},
\be\label{eq:cToThetalm}
\Delta c_{(e),lm}=\sqrt{\frac{(l+2)!}{2(l-2)!}}\Delta \Theta_{lm}\,.
\ee
The above equation will be necessary when we construct the waveform from $\Delta \Theta_{lm}$. 
Specifically, we can compute the waveform by combining  Eqs.~\eqref{eq:Hlm},~\eqref{eq:clmtoUV}, and~\eqref{eq:cToThetalm} in Eq.~\eqref{eq:H} to obtain
\be\label{eq:HtoThetalm}
\Delta h^\mathrm{(disp)}=\frac{1}{\sqrt{2}R}\sum_{l,m}\sqrt{\frac{(l+2)!}{2(l-2)!}}\Delta\Theta_{lm}\,\prescript{}{-2}{Y}_{l m}\,.
\ee
We will denote the memory waveform $\Delta h^\mathrm{(disp)}$ as a sum of the tensor-sourced, $\Delta h^\mathrm{(disp,T)}$, and scalar-sourced $\Delta h^\mathrm{(disp,S)}$ contributions as follows:
\begin{equation} \label{eq:hdispST}
    \Delta h^\mathrm{(disp)} = \Delta h^\mathrm{(disp,T)} + \Delta h^\mathrm{(disp,S)} \, .
\end{equation}
We first compute $\Delta h^\mathrm{(disp,T)}$ followed by $\Delta h^\mathrm{(disp,S)}$.


\subsubsection{Displacement memory effect from the energy flux of tensor GWs} \label{subsubsec:displacementmemorytensor}

The expression for the ``moments'' of $\Delta \Theta^\mathrm{T}$ with respect to $\alpha(x^C)$ have the same general form as in GR, 
\begin{align} \label{eq:ThetaT}
&\int \! d^2\Omega \, \alpha(x^C) \ETH^2 (\ETH^2+2) \Delta \Theta^\mathrm{T} \nonumber \\
& \qquad \qquad \qquad = \int_{u_i}^{u_f} \! du \int d^2\Omega \, \alpha(x^C) N_{AB} N^{AB} \, ,
\end{align}
but there is a subtlety related to the limits of integration ($u_i$ and $u_f$) in the retarded-time integral over $u$.
Because we work in an approximation in which $\xi \ll x$, the lower limit $u_i$ must start at a PN parameter $x_i$ for which $x_i \gg \xi$.
This differs from the corresponding convention in GR, in the limit  $u_i \rightarrow -\infty$ is often taken (in which it is assumed that $x_i \rightarrow 0$). 
The upper limit, $u_f$ is a retarded time at which the corresponding PN parameter $x_f$, is sufficiently large that the PN approximation (at the order at which we work) starts to require higher-PN-order terms to remain accurate. 

The multipolar expansion of $\Delta\Theta^\mathrm{T}$ proceeds exactly as in GR (and we note just a few features of the calculation here; see~\cite{Nichols:2017rqr} for further details).
We can first replace the function $\alpha(x^C)$ with the complex conjugate of a scalar spherical harmonic, $\bar Y_{lm}$.
We then use Eqs.~\eqref{eq:NABfromCAB}, \eqref{eq:cToUV}, \eqref{eq:TABto2Ylm},  and~\eqref{eq:ThetaExpansion}, and the expression for the moments $\Delta \Theta_{l m}^\mathrm{T}$ in terms of the radiative moments $\dot U_{lm}$ and $\dot V_{lm}$ is the same as that derived in GR in~\cite{Nichols:2017rqr}:
\ba\label{eq:DeltaThetalm}
\Delta \Theta_{l m}^\mathrm{T} &=& \frac{1}{2} \frac{(l-2) !}{(l+2) !} \sum_{l^{\prime}, l^{\prime \prime}, m^{\prime}, m^{\prime \prime}} \mathcal{B}_{l}\left(-2, l^{\prime}, m^{\prime} ; 2, l^{\prime \prime}, m^{\prime \prime}\right) \nonumber\\
&& \times \int_{u_i}^{u_{f}} d u\left[ s^{l,(+)}_{l';l''} \left(\dot{U}_{l^{\prime} m^{\prime}} \dot{U}_{l^{\prime \prime} m^{\prime \prime}} +\dot{V}_{l^{\prime} m^{\prime}} \dot{V}_{l^{\prime \prime} m^{\prime \prime}}\right)\right.\nonumber\\
&&\left. \qquad \qquad \quad + 2 i s^{l,(-)}_{l';l''} \dot{U}_{l^{\prime} m^{\prime}} \dot{V}_{l^{\prime \prime} m^{\prime \prime}}\right] \, . 
\ea
We however, introduced the coefficients
\begin{equation} \label{eq:slpm}
    s^{l,(\pm)}_{l';l''} = 1 \pm (-1)^{l+l'+l''}
\end{equation}
that were used in~\cite{Nichols:2018qac} to make the notation more compact.
As in~\cite{Nichols:2017rqr}, the sum runs over $l',l''\geq 2$ and $l$ must be in the range $|l'-l''|\leq l \leq |l'+l''|$ so that the coefficients $\mathcal B_l(-2,l',m';2,l'',m'')$ given in Eq.~\eqref{eq:Bl} are nonzero. 
The azimuthal indices must be related by $m = m'+m''$ for the coefficients $\mathcal B_l(-2,l',m';2,l'',m'')$ to be nonzero.
Because we focus on the leading GW memory effects in the nonoscillatory ($m=0$) part of the waveform, this will further restrict $m'$ and $m''$ to have equal magnitudes and opposite signs: $m'=-m''$. 
While the abstract expression for $\Delta \Theta_{l m}^\mathrm{T}$ in terms of radiative multipole moments has exactly the same form as that in GR, the time-derivatives of the radiative multipole moments $\dot U_{l'm'}$ and $\dot V_{l'm'}$ in BD theory differ from the corresponding moments in GR. 
This leads to a number of order $\xi$ terms in the expression for the GW memory effect that we will compute below.

Next, we will summarize how we compute the memory waveforms, including which radiative multipoles we need and at what PN-order accuracy we require these multipole moments.
For concreteness, let us first focus on products of the mass moments $\dot U_{l'm'} \dot U_{l''m''}$ in Eq.~\eqref{eq:DeltaThetalm}; the arguments will apply to products of current moments and to products of mass and current moments. 
We perform the integral over $u$ by using the chain rule to recast the integral over $u$ in terms of an integral over $x$
\be 
\label{eq:utox_integ}
\int du \dot U_{l'm'} \dot U_{l''m''}=\int \frac{d}{dx}U_{l'm'} \frac {d}{dx}U_{l''m''} \dot x dx\, ,
\ee
as was outlined in, e.g.,~\cite{Favata:2008yd}.
In GR, the Newtonian-order memory waveform can be calculated from just $U_{l'm'}=U_{22}$ and $U_{l''m''}=U_{2,-2}$ (and similarly $U_{l'm'}=U_{2,-2}$ and $U_{l''m''}=U_{22}$), with $U_{22}$ evaluated at Newtonian order, as well. 
In BD theory, however, both $\dot x$ and $d\phi/dx$ (the latter term coming from $dU_{lm}/dx$) have contributions from the dipole moments of the scalar field, and these effects enter at $-1$PN order relative to the GR result (and they are proportional to $\xi$).
To obtain the full result at the Newtonian order requires the 1PN contributions to $U_{22}$, $\dot x$ and $d\phi/dx$. 
Because the $-1$PN terms of $\dot x$ and $d\phi/dx$ are proportional to $\xi$, we only need the parts of the 1PN terms in $U_{22}$, $\dot x$ and $d\phi/dx$ that are independent of $\xi$ to compute $\Delta \Theta_{lm}^\mathrm{T}$. 
Hence, the BD corrections to the 1PN terms in Eqs.~\eqref{eq:U22}, \eqref{eq:fdot} and~\eqref{eq:phase} were not given. 
The $-1$PN term in $\dot x$ also requires that we compute the part of the GW memory waveform sourced by products of the other radiative multipole moments in Eq.~\eqref{eq:radiativeUVmoments} to obtain a Newtonian-order-accurate result; in GR, these moments all give rise to higher-PN corrections to the GW memory effect.

Considering the radiative multipoles described above, we can then compute the multipole moments $\Delta \Theta_{l0}^\mathrm{T}$ of the GW memory effect from Eq.~\eqref{eq:DeltaThetalm}. 
Because the memory is electric-type, only even $l$ moments are nonvanishing for nonspinning compact binaries.
When written in terms of the relevant radiative moments $U_{lm}$ and $V_{lm}$, the moments $\Delta \Theta_{l0}^\mathrm{T}$ are given by
\allowdisplaybreaks
\bw
\begin{subequations}
\begin{align}\label{eq:ThetaHarmonicComponents1}
\Delta \Theta_{20}^\mathrm{T} = {} & \frac{1}{168} \sqrt{\frac{5}{\pi}} \int_{u_i}^{u_f} du \left( 2|\dot U_{22}|^2 - |\dot{V}_{21}|^{2} + \sqrt{7} \Im\left[\sqrt{2}\dot{ \bar U}_{31}\dot V_{21} + \sqrt{5} \dot{ \bar U}_{22}\dot V_{32} \right] + \sqrt 5 \Re\left[\dot{ \bar U}_{42}\dot U_{22} \right] \right) \, , \\ 
\Delta \Theta_{40}^\mathrm{T} = {} & \frac{1}{23760\sqrt{\pi}}  \int_{u_i}^{u_f} du \left( \frac{11}7 |\dot U_{22}|^2  - \frac{44}{7} |\dot{V}_{21}|^{2}  - 21 |\dot U_{33}|^{2} - 7 |\dot U_{31}|^2  + \frac{324\sqrt 5}{7} \Re\left[\dot{ \bar U}_{42}\dot U_{22}\right] \right. \nonumber \\
& \qquad \qquad \qquad \quad \left. + \frac{22}{\sqrt{7}} \Im\left[2\sqrt 5 \dot{ \bar U}_{22}\dot V_{32} + 5 \sqrt{2} \dot{ \bar V}_{21}\dot U_{31} \right] \right) \,,\\
\Delta \Theta_{60}^\mathrm{T} = & -\frac{1}{36960 \sqrt{13\pi}} \int_{u_i}^{u_f} du \left( |\dot U_{33}|^{2} + 15 |\dot U_{31}|^2 - 4\sqrt{5} \Re\left[\dot{ \bar U}_{42}\dot U_{22}\right] \right) \,.
\label{eq:ThetaHarmonicComponents2}
\end{align}
\end{subequations}
\ew
Next, we use Eqs.~\eqref{eq:U22}, \eqref{eq:V21}--\eqref{eq:U42}, and~\eqref{eq:fdot}--\eqref{eq:phase} to perform the integral over $u$ and to write the moments in terms of $x$.
With the identity $(\delta m/M)^2 = 1-4\eta$, we can write the moments as
\begin{subequations}
\label{eq:DeltaThetal0x}
\begin{align}
    \Delta \Theta_{20}^\mathrm{T} = {} & \frac{2\sqrt{5\pi}}{21} M\eta \Delta x \left\{1-\frac 43 \mathscr G -\frac{5\xi S^2}{48 \Delta x}\ln\left(\frac{x_f}{x_i} \right) \right. \nonumber \\
    & \left. -\xi\left[1 + \mathscr F + S^2\left(\frac{1915}{96768}+\frac{665\eta}{1152}\right) \right] \right\} \, , \\
    \Delta \Theta_{40}^\mathrm{T} = {} & \frac{\sqrt{\pi}}{1890} M\eta \Delta x \left\{1 - \frac 43 \mathscr G -\frac{5\xi S^2}{48 \Delta x}\ln\left(\frac{x_f}{x_i} \right) \right. \nonumber \\
    & \left. -\xi\left[1 + \mathscr F + S^2\left(-\frac{737045}{709632}+\frac{143395\eta}{25344}\right) \right] \right\} \, , \\
    \Delta \Theta_{60}^\mathrm{T} = {} & -\sqrt{\frac\pi{13}} \frac{5M \eta S^2 \xi \Delta x}{178827264} (-839+3612\eta) \, .
\end{align}
\end{subequations}
We use the notation $\Delta x=x_f-x_i$, where $x_i$ and $x_f$ correspond to the PN parameter $x$ evaluated at an early retarded time $u_i$ and a final time $u_f$ during the inspiral, respectively.
The terms outside the curly braces in Eq.~\eqref{eq:DeltaThetal0x} in the expressions for $\Delta \Theta_{20}^\mathrm{T}$ and $\Delta \Theta_{40}^\mathrm{T}$ are equal to the equivalent results in GR.
To simplify the notation, we do not include the PN error terms in Eqs.~\eqref{eq:DeltaThetal0x} or~\eqref{eq:Waveform_Disp_Tensor}, which are both are $O[\Delta(x^{3/2})]$, where $\Delta(x^{3/2}) = x_f^{3/2} - x_i^{3/2}$.

Finally, we will construct the displacement memory waveform from the $\Delta \Theta_{l0}$ in Eq.~\eqref{eq:DeltaThetal0x}. 
To do this, it is helpful to have the expressions for the spin-weighted spherical harmonics
\begin{subequations}
\label{eq:m2Yl0}
\begin{align}
    {}_{-2}Y_{20}(\iota,\varphi) = {} & \frac 14 \sqrt{\frac{15}{2\pi}}\sin^2\iota \, ,\\
    {}_{-2}Y_{40}(\iota,\varphi) = {} & \frac 3{16} \sqrt{\frac{10}{\pi}}\sin^2\iota(7\cos^2\iota -1) \, ,\\
    {}_{-2}Y_{60}(\iota,\varphi) = {} & \frac 1{64} \sqrt{\frac{1365}{\pi}}\sin^2\iota(33\cos^4\iota -18\cos^2\iota +1) \, .
\end{align}
\end{subequations}
Substituting Eqs.~\eqref{eq:DeltaThetal0x} and~\eqref{eq:m2Yl0} into Eq.~\eqref{eq:HtoThetalm}, we obtain the displacement memory waveform due to the energy flux from tensor GWs in BD theory.
We find the waveform only contains the plus polarization, and at Newtonian order, it is given by
\begin{widetext}
\begin{align}
\label{eq:Waveform_Disp_Tensor}
    \Delta h_+^\mathrm{(disp,T)} = {} & \frac{\eta  M \Delta x}{48R} \sin^2 \iota (17+\cos^2\iota)\left[ 1 - \frac 43\mathscr G  -\frac{5 \xi  S^2 }{48 \Delta x}\ln\left(\frac{x_f}{x_i}\right) - \xi(1+\mathscr F) - \xi S^2 \left(\frac{81145}{73728} - \frac{65465}{18432}\eta\right)\right. \nonumber \\
    & \left. + \xi S^2 \left(\frac{20975}{172032} - \frac{1075}{2048}\eta \right)\cos^2\iota \right] + \frac{M\eta S^2 \xi \Delta x}{R}\sin^2\iota \left(\frac{783875}{2064384} - \frac{35575}{24576}\eta \right) \, .
\end{align}
\end{widetext}
The expression in front of the square brackets on the first line of Eq.~\eqref{eq:Waveform_Disp_Tensor} is the same as the Newtonian-order waveform for the memory effect in GR.
The terms within the square bracket highlight a number of corrections introduced into the memory waveform amplitude in BD theory.
These include effects related to the change in the amplitude of the $l=2$, $m=\pm 2$ modes (the $\xi$ and $\mathscr G$ terms) and changes in the frequency evolution (the $\mathscr F$ term).
In particular, there is a change in the scaling of the memory with the PN parameter that is proportional to $\ln(x_f/x_i)$, which arises because of scalar dipole radiation.
At the end of the first and on the second line of Eq.~\eqref{eq:Waveform_Disp_Tensor} are a number of terms arising from 1PN-order products of multipole moments coupling to the $-1$PN term in the frequency evolution (or $\dot x$); the terms on the second line lead to a small (order $\xi$) difference to the sky pattern of the memory effect between BD theory and GR.


\subsubsection{Displacement memory effect from the energy flux of scalar radiation}
\label{subsubsec:displacementmemoryscalar}

The displacement memory effect also has a contribution from the integral of the energy flux of the scalar radiation.
Its effect can be computed from the terms proportional to $(\partial_u \lambda_1)^2$ in Eq.~\eqref{eq:DispMemoryFlux}:
\begin{align}
    & \int d^2\Omega \, \alpha(x^C) \ETH^2 (\ETH^2+2) \Delta \Theta^\mathrm{S} \nonumber \\
    & \qquad \qquad \qquad = {}   \frac{6+4\omega}{(\lambda_0)^2} \int_{u_i}^{u_f} \! du \int d^2\Omega \, \alpha(x^C) (\partial_u\lambda_1)^2 \, .
\end{align}
Expanding $\lambda_1$ and $\Delta\Theta^\mathrm{S}$ in scalar spherical harmonics as in Eqs.~\eqref{eq:Lambda_Expansion} and~\eqref{eq:ThetaExpansion}, respectively, and choosing $\alpha = \bar Y_{lm}$, we can determine the multipole moments $\Delta\Theta^\mathrm{S}_{lm}$ in terms of the multipole moments $\lambda_{1(lm)}$ and the integrals of three spherical harmonics defined in Eq.~\eqref{eq:sYlmToBl}.
The result is
\ba\label{eq:ThetaToLambdalm}
\Delta\Theta_{lm}^\mathrm{S} = \frac{(l-2)!}{(l+2)!} \frac{6+4\omega}{(\lambda_0)^2}&&\sum_{l',m',l'',m''}B_l(0,l',m';0,l'',m'')\nonumber\\
&&\times \int_{u_i}^{u_f} du \, \dot \lambda_{1(l'm')} \dot \lambda_{1(l''m'')}\,, \nonumber \\
\ea
where $l\geq 2$ and $l',l''\geq 1$ (and $m$, $m'$, and $m''$ must satisfy $m=m'+m''$). 
Because $\lambda_1$ is proportional to $\xi$, one might be concerned that $\Delta\Theta^\mathrm{S}$ will be an $O(\xi^2)$ effect and be negligible in our approximation.
Note, however, that $3+2\omega = 2/\xi -1$,
which implies that $\Delta\Theta^\mathrm{S}$ is an $\mathcal O(\xi)$ effect; thus, the integrand can be one order higher in $\xi$ and still produce an effect at linear order in $\xi$.

We will now discuss which scalar multipole moments contribute to the displacement memory waveform and the accuracies at which we need the different moments to obtain a Newtonian-order-accurate GW memory waveform, at linear order in $\xi$. 
The 1PN scalar multipole moments $\lambda_{1(lm)}$ that are computed from Eq.~\eqref{eq:scalar_field} are at least $\mathcal O(\xi)$; thus, to linear order in $\xi$, we can treat $6+4\omega$ as $4/\xi$.
We will evaluate the integral over $u$ in Eq.~\eqref{eq:ThetaToLambdalm} by converting it to an integral over $x$ (as was done in Sec.~\ref{subsubsec:displacementmemorytensor}), but unlike in Sec.~\ref{subsubsec:displacementmemorytensor} we need to keep only the GR contribution in $\dot x$, which scales as $x^5$. 
Similarly, when computing $d\phi/dx$, we again need to retain just the GR contribution that goes as $x^{-7/2}$.
The scalar field has a radiative dipole moment, which from Eq.~\eqref{eq:lambda111}, goes as $x^{-1/2}$.
The leading-order part of the integrand (proportional to $\dot\lambda_{1(11)}\dot\lambda_{1(1,-1)}$) scales as $1/x$ rather than $O(x^0)$ as in GR; in this sense, the integrand is a $-1$PN order.\footnote{Note, however, when the integrand is integrated, it will again give rise to a logarithm in $x$ rather than being proportional to $x$, as in GR. 
We will refer to this effect sometimes as a $-1$PN term, since it comes from such an effect in the integrand, and since log terms do not enter into the PN order counting of a term.} 
This product of dipole terms will also contribute to the waveform at 0PN order because of the $O(x^{3/2})$ terms in $\lambda_{1(11)}$; see Eq.~\eqref{eq:lambda111}. 
To work to linear order in $\xi$, we do not need to go to a higher PN order for $\lambda_{1(11)}$.
Similar arguments show that the remaining scalar moments in Eq.~\eqref{eq:scalarmoments} (namely, $\lambda_{1(22)}$ and $\lambda_{1(31)}$) are the ones that are needed to compute Newtonian-order accurate moments of $\Delta \Theta_{l0}^\mathrm{S}$.

We then first list the integrals of the relevant moments $\lambda_{1(lm)}$ that contribute to $\Delta\Theta^\mathrm{S}_{l0}$ at Newtonian order: 
\begin{subequations}
\label{eq:DeltaThetaSl0}
\begin{align}
    \Delta \Theta_{20}^\mathrm{S} = & -\frac{1}{42\sqrt{5\pi}\lambda_0^2 \xi} \int_{u_i}^{u_f} du \bigg( 7|\dot\lambda_{1(11)}|^2 + 10 |\dot\lambda_{1(22)}|^2 \nonumber \\
    & \qquad \qquad - 6\sqrt{14}  \Re\left[\dot \lambda_{1(11)} \dot{\bar{\lambda}}_{1(31)}\right] \bigg) \, , \\
    \Delta \Theta_{40}^\mathrm{S} = {} & \frac{1}{630\sqrt{\pi} \lambda_0^2 \xi} \int_{u_i}^{u_f} du \bigg( |\dot\lambda_{1(22)}|^2 \nonumber \\
    & \qquad \qquad -2 \sqrt{14} \Re\left[\dot \lambda_{1(11)} \dot{\bar{\lambda}}_{1(31)}\right] \bigg) \, .
\end{align}
\end{subequations}
As we did with the the memory sourced by the tensor energy flux, we first substitute the expressions for the scalar moments in Eq.~\eqref{eq:scalarmoments} into Eq.~\eqref{eq:DeltaThetaSl0} and evaluate the integrals in terms of $x$ by using Eqs.~\eqref{eq:fdot}--\eqref{eq:phase}.
This gives the following results:
\begin{subequations}
\label{eq:DeltaThetaSl0x}
\begin{align}
    \Delta \Theta_{20}^\mathrm{S} = & -\frac{M\eta \xi \sqrt{5\pi}}{144} \left\{ S^2\ln\left(\frac{x_f}{x_i}\right)  \right. \nonumber \\
    & \left. + \Delta x\left[\frac 87 \Gamma^2 - \frac{23}{14} S\Gamma \frac{\delta m}M + S^2\left(\frac{71}{336}-\frac{157}{84}\eta\right)\right] \right\} \, , \\
    \Delta \Theta_{40}^\mathrm{S} = {} & \frac{M\eta \xi \sqrt{\pi}\Delta x}{30240} \left(8 \Gamma^2 - \Gamma \frac{\delta m}M + 2 S^2\eta \right) \, .
\end{align}
\end{subequations}
The PN error terms in Eqs.~\eqref{eq:DeltaThetaSl0x} or~\eqref{eq:hdispSplus} are again $O[\Delta (x^{3/2})]$, which we dropped, to simplify the expressions.

We then substitute~\eqref{eq:DeltaThetaSl0x} into Eq.~\eqref{eq:HtoThetalm}, and with the expressions for the spin-weighted spherical harmonics in Eq.~\eqref{eq:m2Yl0}, we arrive at the following equation for the displacement memory waveform sourced by the scalar energy flux:
\begin{align} \label{eq:hdispSplus}
    \Delta h_+^\mathrm{(disp,S)} = & -\frac{5 \eta  M \xi }{192 R} \sin^2\iota \left\{ S^2\ln\left(\frac{x_f}{x_i}\right) \right. \nonumber \\
    & + \Delta x \left[\frac 65 \Gamma^2 - \frac{33}{20} S\Gamma \frac{\delta m}M \right. \nonumber \\
    & + S^2\left( \frac{71}{336} - \frac{113}{60}\eta\right) \nonumber \\
    & \left. \left. - \frac 1{20} \left(8\Gamma^2 - S\Gamma\frac{\delta m}M -2 S^2\eta \right) \cos^2\iota \right] \right\}\,.
\end{align}
There are terms in Eq.~(7.2e) of~\cite{Lang:2013fna} which, after performing the integration over time in our approximation, produce a $-1$PN term in the waveform; this term agrees with the first line of Eq.~\eqref{eq:hdispSplus}.
The Newtonian-order terms require going to a higher PN order than was computed in~\cite{Lang:2013fna}, but the BMS balance laws allow us to determine these expressions.

Because the total GW memory $\Delta h^\mathrm{disp}$ is a sum of the scalar and tensor contributions, as given in Eq.~\eqref{eq:hdispST}, then the scalar-sourced contribution will produce an additional correction to the amplitude and the sky pattern beyond the corrections given in Eq.~\eqref{eq:Waveform_Disp_Tensor} for the tensor-sourced part of the memory effect.

\subsection{GW spin memory effect} \label{subsec:spinmemory}

The GW spin memory effect is a lasting offset in the time integral of the magnetic part of the shear tensor.
It can be constrained through the evolution equation for the Bondi angular momentum aspect, or equivalently the magnetic part of the flux of the super Lorentz charges.
To compute the spin memory effect, it is helpful to denote the time integral of the potential $\Psi$, which gives rise to the magnetic part of $c_{AB}$ in Eq.~\eqref{eq:cToThetaPsi}, as $\Delta \Sigma$:
\be \label{eq:DeltaSigmaPsi}
\Delta\Sigma=\int \Psi du\,.
\ee
We leave off the limits of integration for convenience, though we will later restore these limits when we compute the spin memory in the PN limit.
The generalized BMS balance law for the super angular momentum was given in~\cite{Tahura:2020vsa}, and analogously to the computation in GR (see, e.g.,~\cite{Flanagan:2015pxa}), a term involving Eq.~\eqref{eq:DeltaSigmaPsi} was needed to ensure that the balance law was satisfied.
The form of the balance law can be written as
\begin{subequations}
\begin{equation}\label{eq:BalanceSpinMemory}
    \int d^2\Omega\,\gamma\ETH^2(\ETH^2 + 2) \Delta \Sigma = -\frac{64\pi}{\lambda_0} \lb \Delta Q_{(\gamma)} + \Delta \mathcal J_{(\gamma)}\rb\,,
\end{equation}
where $Y^A=\epsilon^{AB}\eth_B\gamma$ is a smooth, magnetic-parity vector field on the 2-sphere, and where we have defined 
\begin{align}
    \Delta \mathcal J_{(\gamma)} = \frac{\lambda_0}{64\pi} & \int du \, d^2\Omega \, \epsilon^{AD}\eth_D\gamma \bigg[ \eth_A (c_{BC} N^{BC}) \nonumber \\
    & + 2 N^{BC} \eth_A c_{BC} - 4 \eth_B (c_{AC} N^{BC}) \nonumber \\
    & + \frac{4\omega+6}{(\lambda_0)^2} (\partial_u \lambda_1 \eth_A \lambda_1 - \lambda_1 \eth_A \partial_u \lambda_1) \bigg] \, ,
    \label{eq:JSpin}
\end{align}
\begin{align}
    \Delta Q_{(\gamma)} =\frac{\lambda_0}{8\pi} & \int  d^2\Omega \, \epsilon^{AD}\eth_D\gamma \left[- 3\Delta L_A\right.\nonumber\\
    &- \left.\frac 1{4\lambda_0}\Delta(c_{AB} \eth^B \lambda_1 - \lambda_1 \eth^B c_{AB})\right] \, . 
    \label{eq:QSpin}
\end{align}
\end{subequations}
The net flux is denoted by $\Delta \mathcal J_{(\gamma)}$, the change in the charges is denoted by $\Delta Q_{(\gamma)}$, and  the left-hand side of Eq.~\eqref{eq:BalanceSpinMemory} (which is related to the spin memory effect) is the additional term required for the balance law to be satisfied.
Analogously to the displacement memory, the contribution of $\Delta Q_{(\gamma)}$ to Eq.~\eqref{eq:BalanceSpinMemory} is referred to as ordinary spin memory, and $\Delta J_{(\gamma)}$ is the null spin memory (which contains a nonlinear part). 
We will focus on the null contribution to Eq.~\eqref{eq:BalanceSpinMemory} here, as we argue in App.~\ref{Sec:Ordinary_Memory} that the ordinary contribution to the spin memory is likely to be a higher PN effect than the null memory is.

As we did with the change in $\Delta\Theta$ related to the displacement memory effect, we will split $\Delta \Sigma$ into a sum of its contributions from the angular momentum flux of tensor and scalar radiation.
We denote these two contributions by
\begin{equation}
    \Delta \Sigma = \Delta\Sigma^\mathrm{T} + \Delta\Sigma^\mathrm{S} \, , 
\end{equation}
and we compute these two contributions separately below.
 
In addition, while the spin-weight-zero quantity $\Delta\Sigma$ is the most convenient quantity to compute from the balance law~\eqref{eq:BalanceSpinMemory}, the shear $c_{AB}$ or strain $h$ are the more commonly used quantities in gravitational waveform modeling and data analysis.
We thus relate $\Delta\Sigma$ to a time integral of the shear; specifically, we denote the change in the time integral of the magnetic-parity part of the shear tensor by
\be\label{eq:cToSigma}
\int c_{AB,(b)}du=\epsilon_{C(A}\eth_{B)}\eth^C\Delta\Sigma \, .
\ee
Expanding $\Delta \Sigma$ in terms of scalar spherical harmonics
\be\label{eq:SigmaExpansion}
\Delta \Sigma=\sum_{l, m} \Delta \Sigma_{l m} Y^{l m}\left(\iota, \varphi\right)
\ee
(with $l\geq 2$ and $-l\leq m \leq l$), then we can relate the multipole moments $\Delta\Sigma_{lm}$ to the time integrals of the radiative current moments $V_{lm}$ via Eqs.~\eqref{eq:cToSigma}, \eqref{eq:TensorSphMagnetic}, and~\eqref{eq:cToUV}.
The result is that 
\be\label{eq:cToSigmalm}
\int V_{lm} du = \sqrt{\frac{(l+2)!}{2(l-2)!}} \Delta \Sigma_{lm}\,.
\ee
Equation~\eqref{eq:cToSigmalm} allows us to compute the time integral of the strain from $\Delta \Sigma_{lm}$.

\subsubsection{Spin memory effect from the angular momentum flux of tensor GWs}

The null part of the spin memory effect in Eqs.~\eqref{eq:BalanceSpinMemory} and~\eqref{eq:JSpin}, that is sourced by the tensor GWs can be computed from the following expression:
\begin{widetext}
\begin{equation} \label{eq:BalanceLawTensorSpin}
\int d^2\Omega\,\gamma\ETH^2(\ETH^2 + 2) \Delta \Sigma^{\mathrm{T}} = \int_{u_i}^{u_f} du \int d^2\Omega \, \epsilon^{AD}\eth_D\gamma \bigg[ \eth_A (c_{BC} N^{BC}) +  2 N^{BC} \eth_A c_{BC} - 4 \eth_B (c_{AC} N^{BC})\bigg]\,.
\end{equation}
\end{widetext}
It has the same form as the analogous expression in GR.
It can then be recast into the same form as in Eq.~(3.23) of~\cite{Flanagan:2015pxa} by using the identities in Appendix~C of~\cite{Flanagan:2015pxa}.
This expression was then the starting point for the multipolar expansion of the spin memory effect given in~\cite{Nichols:2017rqr}.
We reproduce the result from~\cite{Nichols:2017rqr} below; however, we first introduce, in addition to $s^{l,(\pm)}_{l';l''}$ in Eq.~\eqref{eq:slpm}, the following coefficients (defined in~\cite{Nichols:2018qac}) to make the expression more concise:
\begin{align}
    c^l_{l'm';l''m''} = {} & 3 \sqrt{\left(l'-1\right)\left(l'+2\right)}  \mathcal{B}_{l}\left(-1, l',m'; 2,l'',m''\right) \nonumber \\
    & + \sqrt{\left(l''-2\right)\left(l''+3\right)}  \mathcal{B}_{l}\left(-2, l', m';  3, l'', m''\right)\,.
\end{align}
The expression for the moments $\Delta\Sigma_{lm}$ can then be derived through a lengthy calculation outlined in~\cite{Nichols:2017rqr}, and the result is given by
\bw
\begin{align} \label{eq:Sigmalm}
    \Delta \Sigma_{l m}^\mathrm{T} = \frac{1}{4 \sqrt{l(l+1)}} \frac{(l-2) !}{(l+2) !} \sum_{\stackrel{l' ,l'',}{m' ,m''}}  c^l_{l'm';l''m''} \int_{u_i}^{u_{f}} \! du & \left[i s^{l,(-)}_{l';l''} \left(U_{l' m'} \dot{U}_{l'' m''}-\dot{U}_{l' m'} U_{l'' m''}+V_{l' m'} \dot{V}_{l'' m''}-\dot{V}_{l' m'} V_{l'' m''}\right) \right. \nonumber \\
    & \left. - s^{l,(+)}_{l';l''} \left(U_{l' m'} \dot{V}_{l'' m''}+\dot{V}_{l' m'} U_{l'' m''}-\dot{U}_{l' m'} V_{l'' m''}-V_{l' m'} \dot{U}_{l'' m''}\right)\right] \, .
\end{align}
\ew

To compute the spin memory effect at Newtonian order, we need precisely the same set of radiative multipole moments $U_{lm}$ and $V_{lm}$ that were used to compute the displacement memory effect in Sec.~\ref{subsubsec:displacementmemorytensor}.
We then compute the spin memory effect following the same procedure in Sec.~\ref{subsubsec:displacementmemorytensor} by first writing the needed moments of $\Delta \Sigma_{lm}$ in terms of integrals of the relevant radiative moments $U_{lm}$ and $V_{lm}$:
\bw
\begin{subequations}
\begin{align}\label{eq:Sigma30}
    \Delta\Sigma_{30}^\mathrm{T} = - \frac{1}{720\sqrt{7\pi}} \int_{u_i}^{u_f} \! du & \Bigg\{ \Im\left[9 \left(\dot{\bar U}_{22}U_{22} - 2\dot{\bar V}_{21}V_{21}\right) + 7 \left(\dot{\bar U}_{31}U_{31} - \dot{\bar U}_{33}U_{33} \right) + 11\sqrt{5} \left(\dot{\bar U}_{22}U_{42} + \dot{\bar U}_{42}U_{22}\right) \right] \nonumber \\
    & \left. + \Re\left[5 \sqrt{35} \left(\dot{\bar V}_{32}U_{22} - \dot{\bar U}_{22}V_{32} \right) - 5\sqrt{\frac 72} \left(\dot{\bar V}_{21}U_{31}-\dot{\bar U}_{31}V_{21}\right) \right] \right\} \, ,\\
    \Delta\Sigma_{50}^\mathrm{T} = \frac{1}{5040\sqrt{11\pi} } \int_{u_i}^{u_f} \! du & \Bigg\{ \Im\left[ 5\left(\dot{\bar U}_{33}U_{33} + 5 \dot{\bar U}_{31}U_{31}\right) - \frac{38}{\sqrt 5} \left(\dot{\bar U}_{22}U_{42} + \dot{\bar U}_{42}U_{22}\right)\right] \nonumber \\
    & \left. + \Re\left[ 2\sqrt{\frac 75} \left(\dot{\bar U}_{22}V_{32}-\dot{\bar V}_{32}U_{22} \right) + 2\sqrt{14} \left(\dot{\bar V}_{21}U_{31}-\dot{\bar U}_{31}V_{21}\right) \right] \right\} \, .
    \label{eq:Sigma50}
\end{align}
\end{subequations}
\ew
As in Sec.~\ref{subsubsec:displacementmemorytensor}, we then use Eqs.~\eqref{eq:U22}, \eqref{eq:V21}--\eqref{eq:U42}, and~\eqref{eq:fdot}--\eqref{eq:phase} to evaluate the integrals in Eqs.~\eqref{eq:Sigma30}--\eqref{eq:Sigma50} and to write the expression for the moments $\Delta\Sigma_{30}$ and $\Delta\Sigma_{50}$ in terms of $x$.
Unlike in Sec.~\ref{subsubsec:displacementmemorytensor}, the integrand does not depend on $\dot x$, when written as an integral over $x$, because there is only one time derivative of the radiative multipole moments.
The result of this integration is given by
\begin{subequations}\label{eq:Sigmainx}
\begin{align} \label{eq:Sigma30inx}
\Delta \Sigma_{30}^\mathrm{T} = {} & \sqrt{\frac \pi 7}\frac{\eta  M^2}{10} \left\{-\frac{5 \xi  S^2}{144}\Delta(x^{-3/2}) + \Delta(x^{-1/2}) \left[1-\frac{4}{3}\mathscr G \right. \right. \nonumber\\
& - \left. \left. (1+\mathscr F)\xi + \frac{5 (47796 \eta +5003) S^2 \xi}{435456}\right]\right\} \, ,\\
\label{eq:Sigma50inx}
\Delta\Sigma_{50}^\mathrm{T} = & - \sqrt{\frac{\pi}{11} } \frac{ \eta  M^2 \xi  S^2}{12192768} (21588 \eta -4117)\Delta(x^{-1/2}) \, ,
\end{align}
\end{subequations}
where we have defined $\Delta(x^{-1/2}) = x_f^{-1/2} - x_i^{-1/2}$ and similarly for $\Delta(x^{-3/2})$.
The PN error terms in Eqs.~\eqref{eq:Sigmainx} and~\eqref{eq:hx_int_spinT} are both expected to be of order $\log(x_f/x_i)$, which would arise from $O(1/x)$ terms in the $u$ integral.
We did not write these terms out explicitly, so as to simplify the notation.

In GR, the only term that appears in the Newtonian-order spin-memory waveform is $\Delta(x^{-1/2})$ times the coefficient outside the curly braces in Eq~\eqref{eq:Sigma30}.
The remaining terms in $\Delta \Sigma_{30}$ and the entire expression for $\Delta \Sigma_{50}$ appear in the BD-theory waveform at Newtonian order because of the $-1$ PN term in $d\phi/dx$ and the additional radiative multipole moments that contribute in BD theory, but not in GR.

Finally, we construct the time-integrated strain from the moments $\Delta \Sigma_{l0}$ in Eqs.~\eqref{eq:Sigma30inx}--\eqref{eq:Sigma50inx}. 
Using Eqs.~\eqref{eq:H}, \eqref{eq:Hlm}, \eqref{eq:clmtoUV}, and~\eqref{eq:cToSigmalm}, we can write the relation between the time integral of $h$ and a general $\Delta \Sigma_{lm}$ as
\be\label{eq:HtoSigmalm}
\int_{u_i}^{u_f} h^\mathrm{(spin)} du = \frac{-i}{2R}\sum_{l,m}\sqrt{\frac{(l+2)!}{(l-2)!}}\Delta\Sigma_{lm}(\prescript{}{-2}{Y}_{l m})\,.
\ee
Because the modes $\Delta\Sigma_{l0}$ that produce the time-integrated strain $h^\mathrm{(spin)}$ in Eq.~\eqref{eq:Sigma30inx} and~\eqref{eq:Sigma50inx} are real, then from  Eq.~\eqref{eq:WaveformPlusCross} it follows that the spin memory enters in the cross mode polarization of gravitational waves (as it does in GR~\cite{Nichols:2017rqr}). 
Finally substituting Eqs.~\eqref{eq:Sigma30inx}--\eqref{eq:Sigma50inx} into~\eqref{eq:HtoSigmalm}, and using the expressions for the spin-weighted spherical harmonics
\begin{subequations}
\begin{align}
    {}_{-2}Y_{30}(\iota,\varphi) = & {} \frac 14 \sqrt{\frac{105}{2\pi}} \sin^2\iota \cos\iota \, , \\
    {}_{-2}Y_{50}(\iota,\varphi) = & {} \frac 18 \sqrt{\frac{1155}{2\pi}} \sin^2\iota \cos\iota(3\cos^2\iota - 1) \, ,
\end{align}
\end{subequations}
we obtain for the time-integrated strain
\bw
\begin{align} \label{eq:hx_int_spinT}
\int^{u_f}_{u_i} \! du \, h_\times^\mathrm{(spin, T)} = {} & \frac{3 \eta  M^2}{8 R}\sin ^2\iota \cos\iota\left\{-\frac{5 \xi  S^2}{144} \Delta(x^{-3/2}) + \left[1-\frac{4}{3} \mathscr G-(1+\mathscr F)\xi- \xi  S^2 \left(\frac{3365}{6912} \eta + \frac{1915}{27648} \right) \right]\Delta(x^{-1/2}) \right.\nonumber\\
& \left.
+ \xi S^2 \left(\frac{20585}{580608} - \frac{1285}{6912} \eta\right)\Delta(x^{-1/2}) \cos^2\iota  \right\} \,.
\end{align}
\ew
The expression for the $u$ integral of $h_\times^\mathrm{(spin, T)}$ is written such that the angular dependence and coefficient $3\eta M^2/(8R)$ outside of the curly braces coincides with the expression in GR at Newtonian order.
Within the curly braces there are several sorts of terms: (i) the first term proportional to $\Delta(x^{-3/2})$ is a $-1$PN term arising from the dipole term in the phase, but which has the same angular dependence as the spin memory effect in GR; (ii) the terms in the square bracket (aside from the factor of unity that reproduces the GR expression for the spin memory) are small BD corrections (proportional to $\xi$) that modify the amplitude of the waveform without changing its $x$ or $\iota$ dependence; (iii) the final terms on the second line are those which have the same $x$ dependence, but a different angular dependence from the GR expression (and are again proportional to $\xi$).

\subsubsection{Spin memory effect from the angular momentum flux of the scalar radiation}

The angular momentum flux from the scalar radiation produces a second contribution to the spin memory effect. 
Its contribution can be obtained from the scalar field terms in the balance law in Eq.~\eqref{eq:BalanceSpinMemory},
\begin{align}
\int d^2\Omega\gamma\ETH^4 (\ETH^2+2) \Delta \Sigma^{\mathrm{S}} = - & \frac{6+4\omega}{(\lambda_0)^2} \int_{u_i}^{u_f} d^2\Omega du \epsilon^{AB}\eth_B\gamma\nonumber\\
& \times(\dot\lambda_1 \eth_A \lambda_1 - \lambda_1 \eth_A \dot \lambda_1) \, .
\end{align}
The multipolar expansion of $\Delta \Sigma$ can be obtained by assuming $\gamma$ is equal to the spherical harmonic $\bar Y_{lm}$, and then using the multipolar expansion of $\lambda_1$ in Eq.~\eqref{eq:Lambda_Expansion}.
After relating the gradients and curls of spherical harmonics in this expansion to the electric- and magnetic-parity vector harmonics in  Eqs.~\eqref{eq:VectorSphElectric}--\eqref{eq:VectorSphMagnetic} and then employing the relationship between these vector harmonics and the spin-weighted spherical harmonics in Eqs.~\eqref{eq:VectorSphElectricDyads}--\eqref{eq:VectorSphMagneticDyads},we can derive the moments $\Delta \Sigma_{lm}$ in terms of moments $\lambda_{1(lm)}$ (and their time derivatives and complex conjugates) and the coefficients $\mathcal B_l(s',l',m';s'',l'',m'')$ given in Eq.~\eqref{eq:sYlmToBl}.
The resulting expression is given below:
\begin{align}
    \Delta \Sigma^{\mathrm{S}}_{lm} = {} & \frac{i(2\omega+3)}{\lambda_0^2\sqrt{l(l+1)}} \frac{(l-2)!}{(l+2)!} \sum_{l',m',l'',m''} s^{l,(-)}_{l';l''} \nonumber\\
    & \times \sqrt{l''(l''+1)} \mathcal B_l(0,l',m';1,l'',m'')\nonumber\\
    &\times\int_{u_i}^{u_f} du(\dot \lambda_{1(l'm')} \lambda_{1(l''m'')}- \lambda_{1(l'm')}\dot\lambda_{1(l''m'')}) \,.
\end{align}
The coefficient $s^{l,(-)}_{l';l''}$ is defined in Eq.~\eqref{eq:slpm}.

The moments of $\lambda_{1(lm)}$ that contribute to the Newtonian-order spin memory effect are the same moments needed to compute the scalar-sourced displacement memory effect in Sec.~\ref{subsubsec:displacementmemoryscalar}.
The only nonvanishing moment of $\Delta\Sigma$ at this order then is  $\Delta \Sigma_{30}$, and to linear order in $\xi$ it is given by
\begin{align}
\label{eq:Sigma30Scalar}
    \Delta\Sigma^{\mathrm{S}}_{30} = - & \frac{1}{30\xi(\lambda_0)^2} \frac 1{\sqrt{7\pi}} \int_{u_i}^{u_f} \! du \, \Im\bigg[\dot \lambda_{1(22)} \bar \lambda_{1(22)} \nonumber\\
     & - \sqrt{\frac 72 } \left(\dot \lambda_{1(11)} \bar \lambda_{1(31)}-\lambda_{1(11)} \dot{\bar \lambda}_{1(31)}\right) \bigg] \, . 
\end{align}
We can then use the expressions for the scalar-field multipole moments in Eq.~\eqref{eq:scalarmoments} to evaluate the integrals and write them in terms of $x$ (analogously to what was done for the tensor-GW part of the spin memory effect), and we find it is given by
\begin{equation} \label{eq:DeltaSigma30S}
    \Delta\Sigma^{\mathrm{S}}_{30} =   \frac{\sqrt{\pi} \eta  M^2 \xi }{1440\sqrt{7}}\left(2 \eta  S^2 + \Gamma S\frac{\delta m}{M} - 8 \Gamma ^2 \right)\Delta\lb x^{-1/2} \rb \, .
\end{equation}
It is then straightforward to use Eq.~\eqref{eq:HtoSigmalm} to write the retarded-time integral of the spin memory waveform generated by scalar angular momentum flux as
\begin{align} \label{eq:Int_hxS}
\int^{u_f}_{u_i} du h_\times^\mathrm{(spin, S)}
   = {} & \frac{\eta  M^2\xi}{384 R}\left(2 \eta  S^2+\Gamma  \frac{\delta m}{M} S -8 \Gamma ^2\right)\nonumber\\
   &\times\Delta\lb x^{-1/2} \rb \sin^2 \iota \cos \iota\,.
\end{align}
The PN error terms in Eqs.~\eqref{eq:DeltaSigma30S} and~\eqref{eq:Int_hxS} are both expected to be of order $\log(x_f/x_i)$, which would arise from $O(1/x)$ terms in the $u$ integral.

The full retarded-time integral of the spin memory waveform is $h_\times^\mathrm{(spin)} = h_\times^\mathrm{(spin, S)} + h_\times^\mathrm{(spin, T)}$.
It then adds small correction linear in $\xi$ to Eq.~\eqref{eq:hx_int_spinT} that changes the amplitude but does not alter the $x$ or $\iota$ dependence of the effect.

\section{Conclusions}\label{sec:conclusions}

In this paper, we computed the displacement and spin GW memory effects generated by nonprecessing, quasicircular binaries in BD theory. 
We worked in the PN approximation, and the expressions that we computed are accurate to leading Newtonian order in the PN parameter $x$, and they include the leading-order corrections in the BD parameter $\xi$. 
Our calculations relied upon using the BMS balance laws associated with the asymptotic symmetries in BD theory (which are the same as in GR).
These balance laws permit us to determine the tensor GW memory effects, but the scalar GW memory effects associated with the breathing polarization are not constrained through these balance laws. 
We further focused on the null contributions to the tensor GW memory effects, because we estimated that the ordinary (linear) memory effects would contribute at a higher PN order than the null (including the nonlinear) memory effects.

Using the BMS balance laws has the advantage that we can determine the memory effects in BD theory to a higher PN order than had been previously done through the direct integration of the relaxed Einstein equations in harmonic gauge.
However, the balance laws take as input radiative and nonradiative data at large Bondi radius and this data must be obtained through some other method.
Specifically, in the context of this paper in PN theory, we needed to take as input the oscilliatory scalar and tensor GWs computed in harmonic gauge in BD theory in~\cite{Lang:2014osa}.
This required us to compute a coordinate transformation between harmonic and Bondi gauges at leading order in the inverse distance to the source, so that we could relate the radiative GW data in these two different coordinate choices (and formalisms). 
There were relatively simple transformations that allowed us to relate the Bondi shear tensor to the transverse-traceless components of GW strain, and these relations were particularly simple when expressed in terms of the radiative multipole moments of the shear and strain tensors. 
There were similar expressions relating the scalar field at leading order in inverse distance in the two coordinate systems and the corresponding multipole moments of the scalar field. 
With these relationships between the multipolar expansions of the scalar field and the shear tensor in harmonic and Bondi gauges, we could then use the BMS balance laws to compute the GW memory effects.

The tensor GW memory effects in BD theory have several noteworthy differences from the corresponding effects in GR at the equivalent PN order. 
First, because of scalar dipole radiation in BD theory, there are relative $-1$PN-order terms in the memory effects in BD theory. 
The $-1$PN term in the displacement memory waveform comes from two sources in the supermomentum balance law: (i) directly from the energy flux of scalar radiation and (ii) indirectly from the energy flux in tensor radiation (specifically through dipole contributions to the frequency evolution and the GW phase). 
The spin memory waveform, however, has a relative $-1$PN correction from GR arising from only the energy flux of the tensor waves (the scalar-sourced part of the energy flux gives rise to a contribution at the same leading order as in GR, and it comes from products of dipole and octupole moments, as well as quadrupole moments with themselves). 
The absence and presence of the $-1$PN term from the scalar radiation in the spin and displacement memory effects, respectively, arises because of the different lowest multipole term in the sky pattern of the effects: the spin memory effect begins with the $l=3$, $m=0$ mode, whereas the displacement memory has an $l=2$, $m=0$ mode. 

A second noteworthy feature is that the computation of Newtonian-order memory effects in BD theory requires radiative multipole moments of the strain at higher PN orders than are required in GR (in which the leading-order effects can be computed from just the $l=2$, $m=\pm 2$ radiative mass quadrupole moments. 
Because there are $-1$PN terms in the GW phase and the evolution of the GW frequency, computing the Newtonian GW memory effects requires higher-PN-order radiative multipole moments (including the current quadrupole, the mass and current octupoles, and the mass hexadecapole). 
These higher-order mass and current multipole moments also give rise to a sky pattern of the GW memory effect in BD theory that differs from the sky pattern of the effect in GR.
In addition, the presence of dipole radiation required us to include $1$PN corrections to the GW phase and frequency time derivative to compute the memory effects to Newtonian order (though we only required the GR limit of these 1PN corrections when working to linear order in the BD parameter $\xi$).

Finally, let us conclude by commenting on potential applications of the calculations of GW memory effects given in this paper. 
Because the calculations herein have shown that the memory effect in BD theory differs from that in GR, it is natural to ask if these differences could be detected.
Given the challenges of detecting the memory effect in GR with LIGO and Virgo~\cite{Hubner:2019sly,Boersma:2020gxx,Hubner:2021amk} and the fact that the PN corrections are small, it would be more natural to consider whether next-generation gravitational-wave detectors---such as the space-based detector LISA~\cite{Audley:2017drz} or ground-based detectors like the Einstein Telescope or Cosmic Explorer~\cite{Maggiore:2019uih,Reitze:2019iox}---could constrain BD theory through a measurement of the GW memory effect.
The constraints could come from searching for differences in the leading-order amplitude of the effect, in the time dependence of the accumulation of the memory effect (through the different dependence on the PN parameter $x$), or in the sky pattern of memory effects in BD theory (the latter being similar to the hypothesis test described in~\cite{Yang:2018ceq}).
Because memory effects accumulate most rapidly during the merger of compact binaries, having waveforms that go beyond the PN approximation and include the merger and ringdown would be important for producing the most accurate constraints.
Nevertheless, we use the PN results to give order-of-magnitude estimates for the prospects of detecting any BD effects in the memory waveforms.

The BD corrections to the GW memory waveforms computed in this paper depend on the BD parameter $\xi$ and the sensitivities of the binary's components. 
The dependence on the sensitivities is such that for binary black holes, which have binary components with sensitivities equal to one half in BD theory, the Newtonian GW memory waveform for binary black holes is equivalent to the waveform in GR after rescaling the black-hole masses by $(1-\xi/2)$.
The fact that there exists a constant redefinition of the mass that produces the same waveform as in GR implies that the difference is not observable, as was previously noted in Refs.~\cite{Will:1989sk,Mirshekari:2013vb,Lang:2013fna} in PN theory.
This also implies that observations of GWs from supermassive binary black holes by the LISA detector, although possibly high in signal to noise~\cite{Audley:2017drz}, may not be able detect or constrain features of BD theory.

Mixed black-hole neutron-star binaries, however, will have observable effects, because the objects have different sensitivities. 
From Eqs.~\eqref{eq:Waveform_Disp_Tensor},~\eqref{eq:hdispSplus},~\eqref{eq:hx_int_spinT}, and~\eqref{eq:Int_hxS}, it follows that the leading PN BD correction is $\mathcal O (\xi/\Delta x)$ smaller than the leading PN GR memory effects.\footnote{The $-1PN$ BD corrections to displacement memory effects are of $\mathcal O ((\xi/\Delta x)\ln[x_f/x_i])$. However, considering $x_i$ to be the PN parameter at lower cut-off frequency of the detector and $x_f$ to be at the innermost stable circular orbit frequency, we expect $\ln[x_f/x_i]$ to be of order unity.}
Given the current lower bound on $\xi$ and taking the final PN parameter $x_f$ to be PN parameter at the frequency of the innermost stable circular orbit, the $-1$PN BD corrections are $\mathcal O (10^{-4})$ times smaller in amplitude than the 0PN GR memory effects.
While we have not performed a detailed signal-to-noise calculation, this does make it appear challenging to detect the effects if present. 
However, it is difficult to provide an accurate estimate for the signal-to-noise of the BD memory effects for the following reasons. 
First, to the best of our knowledge, there has not been a systematic study of the  signal to noise for GW memory effects arising from black-hole neutron-star binaries, even in GR. 
Second, memory effects accumulate most rapidly during the merger of compact binaries, during which the PN approximation becomes less valid and numerical simulations of the merger and ringdown are required to produce accurate waveforms.
We thus leave a quantitative assessment of the detection prospects of the GW memory effects in BD theory for future work.


\acknowledgments

S.T.\ and K.Y.\ acknowledge support from the Owens Family Foundation. K.Y.\ also acknowledges support from the NSF Grant No.~PHY-1806776, the NASA Grant 80NSSC20K0523, and a Sloan Foundation Research Fellowship. 
K.Y.\ would like to also thank the support by the COST Action GWverse CA16104 and JSPS KAKENHI Grants No. JP17H06358.
D.A.N.\ acknowledges support from the NSF Grant No.~PHY-2011784 and the ORAU Ralph E.~Powe Junior Faculty Enhancement Award.

\appendix

\section{Coordinate transformations from harmonic gauge to Bondi gauge}\label{app1}

In this appendix, we construct a coordinate transformation between harmonic coordinates to linear order in $1/R$ and Bondi coordinates at an equivalent order in $1/r$ for a radiating spacetime in Brans-Dicke theory. 
The procedure is similar to that recently described in~\cite{Blanchet:2020ngx} in GR, but we do not work to all orders in $1/r$ as in~\cite{Blanchet:2020ngx}.
Rather, we work only to an order in $1/r$ so that we can relate the radiative data in harmonic gauge ($\tilde h_{ij}^\mathrm{TT}$ and $\Xi$) to the corresponding radiative data in Bondi gauge ($c_{AB}$ and $\lambda_1$) and compute the GW memory effects from PN waveforms in harmonic gauge.

We will denote the harmonic-gauge metric by $g^\mathrm{(H)}_{\mu\nu}$ which we write in quasi-Cartesian coordinates $X^\mu = (X^0,X^i)$, where $X^0 = t$ and $X^i=(X,Y,Z)$.
We find it convenient to define $R = \sqrt{X^i X^j \delta_{ij}}$ to be the harmonic-gauge distance from the origin, and $y^A = (\iota,\varphi)$ to be spherical polar coordinates with $\cos\iota = Z/R$ and $\tan\varphi = Y/X$.
The components of the metric can then be written in the form 
\allowdisplaybreaks
\begin{subequations}
\label{eq:MetricHarmonic}
\ba\label{eq:MetricHarmonic1}
g^{(H)}_{00}&=&-1+\frac{2AM}{R}+\frac{1}{R}H_{00}(t-R,y^A)\,,\\
g^{(H)}_{0i}&=&\frac{1}{R}H_{0i}(t-R,y^A)\,,\\
g^{(H)}_{ij}&=&\delta_{ij}+\frac{2BM}{R}\delta_{ij}+\frac{1}{R}H_{ij}(t-R,y^A) \label{eq:MetricHarmonic3} \, .
\ea
\end{subequations}
We have written the metric in the center-of-mass rest frame of the system, and we have split the $O(1/R)$ part of the metric in terms of the constant mass monopole moment $M$ and all the time-dependent, higher-order multipole moments in $H_{\mu\nu}$. 
We have also introduced constants $A$ and $B$ [defined in Eq.~\eqref{eq:staticBDparams}] which are needed so that the metric satisfies the modified Einstein's equations in the static limit (at the leading nontrivial order in $1/R$).
To specify a solution, we also need an expression for the scalar field, which we give in the Jordan frame, and which we denote by $\lambda$.
We will again expand it in terms of a static monopole moment and time-dependent, higher-order multipole moments as follows: 
\be \label{eq:ScalarField}
\lambda=\lambda_0 \left[1+\frac{\psi(t-R,y^A)+2C M}{R} \right]\,.
\ee 
The monopole term is the $O(1/R)$ piece proportional to $2\lambda_0 C M$ and $\psi$ contains the higher-order, time-dependent moments. 
The third coefficient $C$ is again needed to satisfy the field equations in the static limit.
The expressions for the constants $A$, $B$, and $C$ were previously determined in~\cite{Lang:2013fna,Will:1989sk,Zaglauer:1992bp} and are given by
\begin{subequations}
\label{eq:staticBDparams}
\ba\label{eq:parameters1}
A &=& \frac{\omega + 2 - \kappa}{\omega +2} \,,\\
B &=& \frac{\omega + 1 + \kappa}{\omega +2} \,,\\
C &=& \frac{1-2 \kappa }{2 \omega +4}\label{eq:parameters3}\, .
\ea
\end{subequations}
We introduced the variable $\kappa=(m_1 s_1+m_2s_2)/M$ in the equations above.
The coefficients also satisfy the relationships $A-B=2C$ and $A+B = 2/\lambda_0$. 

The harmonic gauge conditions lead to relationships between the components of the quantities $H_{\mu\nu}$ and $\psi$. 
These relationships are more conveniently expressed in terms of a quantity $\tilde H_{\mu\nu}$, which is related to $H_{\mu\nu}$ and $\psi$ by
\ba\label{eq:htildeToh}
H_{\mu\nu}=\tilde H_{\mu\nu}-\frac{1}{2}\tilde H \eta_{\mu\nu}-\psi \eta_{\mu\nu} \, .
\ea
The gauge conditions are given by 
\ba \label{eq:HarmonicGauge}
\tilde H_{00}=n^i n^j \tilde H_{ij}\,,\qquad \tilde H_{0i}=-n^j\tilde H_{ij}\, .
\ea
We have defined $n^i = X^i/R = \partial_i R$ above, and it reduces to the expression $\vec n\equiv(\sin\iota \cos \varphi,\sin \iota \sin \varphi,\cos \iota)$ when written in terms of $\iota$ and $\varphi$.\footnote{The conditions in Eq.~\eqref{eq:HarmonicGauge} can be derived by first making the definitions given, e.g., in~\cite{Will:1989sk,Mirshekari:2013vb}, of a conformally rescaled metric $\tilde g_{\mu\nu} = \lambda g_{\mu\nu}$, then defining $\tilde h^{\mu\nu}$ to be $\tilde h^{\mu\nu} = \eta^{\mu\nu} - \sqrt{-\tilde g} \tilde g^{\mu\nu}$, and imposing the harmonic gauge condition $\partial_{\nu} \tilde h^{\mu\nu}=0$. 
When the harmonic gauge condition is imposed at leading order in $1/R$, then the conditions in~\eqref{eq:HarmonicGauge} can be obtained up to integration constants that we set to zero [so as to maintain the static solution of Einstein's equations in Eqs.~\eqref{eq:MetricHarmonic} and~\eqref{eq:ScalarField}].}

Our procedure for transforming from harmonic gauge to Bondi gauge follows some aspects of the work~\cite{Blanchet:2020ngx} in GR, in which Bondi coordinates were determined in terms as functions of harmonic-gauge quantities. 
In our case, however, we consider BD theory, we work only to linear order in $1/R$ in harmonic coordinates, and we determine the corresponding Bondi coordinates in an expansion in $1/R$, such that Bondi gauge is imposed to one order in $1/r$ beyond the leading-order metric. 
To perform the coordinate transformation, it is useful to work with the components of the inverse metric.
In harmonic gauge, these components are given by
\begin{equation}
    g^{\mu\nu}_\mathrm{(H)} = \eta^{\mu\nu}_\mathrm{(H)} - \frac 1R \left[ H^{\mu\nu} + 2M (A \delta^\mu_0 \delta^\nu_0 + B \delta^\mu_i \delta^\nu_j \delta^{ij}) \right] + O(R^{-2}) \, ,
\end{equation}
where $H^{\mu\nu}$ is related to $H_{\mu\nu}$ by raising indices with $\eta^{\mu\nu}_\mathrm{(H)}$.
We aim to put the metric in Bondi form, in which the nonzero components of the inverse Bondi metric are given by
\begin{subequations}
\label{eq:InverseBondiMetric}
\begin{align}
g_\mathrm{(B)}^{ur} = & -1 -\frac{\lambda_1}{\lambda_0 r} + O\lb r^{-2} \rb \, , \\
g_\mathrm{(B)}^{rr} = {} & 1 + \frac{\partial_u\lambda_1}{\lambda_0}+\frac{1}{r}\left(\frac{\lambda_1}{\lambda_0}-2\mathcal M+\frac{\lambda_1\partial_u\lambda_1}{2\lambda_0^2}\right)\nonumber\\
& +\mathcal O  \lb r^{-2}\rb \, , \\
g_\mathrm{(B)}^{rA}  = {} & \frac{1}{2r^2} \lb \eth_B c^{AB} - \frac{\eth^A\lambda_1}{\lambda_0}\rb + O\lb r^{-3}\rb \, , \\
g_\mathrm{(B)}^{AB}  = {} & \frac 1{r^2} q^{AB} - \frac 1{r^3} c^{AB} + O(r^{-4} )\,
\end{align}
\end{subequations}
(and where $g_\mathrm{(B)}^{uu} = g_\mathrm{(B)}^{uA} = 0$).
For simplicity, we will summarize Eq.~\eqref{eq:InverseBondiMetric} as
\begin{equation}
    g_\mathrm{(B)}^{\mu\nu} =  \eta_\mathrm{(B)}^{\mu\nu} - \frac 1r  h_\mathrm{(B)}^{\mu\nu} \, ,
\end{equation}
where the quantity $\eta_\mathrm{(B)}^{\mu\nu}$ consists of the $O(r^0)$ pieces of $g_\mathrm{(B)}^{ur}$ and $g_\mathrm{(B)}^{rr}$, the $O(r^{-1})$ part of $g_\mathrm{(B)}^{rA}$ (which vanishes) and the $O(r^{-2})$ part of $g_\mathrm{(B)}^{AB}$; the quantity $h_\mathrm{(B)}^{\mu\nu}$ consists of the coefficients of the relative $1/r$ corrections to the components of $\eta_\mathrm{(B)}^{\mu\nu}$.

We perform the coordinate transformation in two stages for illustrative purposes (one could perform it in one stage as in~\cite{Blanchet:2020ngx}, but the two-stage process here allows us to highlight the different roles of the different terms in the transformation more easily).
The first stage imposes the gauge conditions on the inverse Bondi metric that $g_\mathrm{(B)}^{uu}$ and $g_\mathrm{(B)}^{uA}$ vanish to the accuracy in $1/r$ at which we work, it also makes $r$ an areal radius, and finally, it relates the Bondi-coordinate retarded time $u$ to the harmonic-gauge retarded time $t-R$.  
It can be thought of as a ``finite'' gauge transformation, in the sense that it is needed to relate the background metrics $\eta_\mathrm{(B)}^{\mu\nu}$ and $\eta_\mathrm{(H)}^{\mu\nu}$, which differ by a large (not perturbative in $1/r$) coordinate transformation.
The second stage, which can be treated as perturbative in $1/r$, then sets the metric following the first transformation into a Bondi-gauge metric that satisfies the modified Einstein equations of BD theory.

The first stage, the finite  part of the coordinate transformation expresses a set of coordinates $x^\alpha = (u,r,x^A)$ in terms of harmonic gauge coordinates $X^\alpha = (t,X^i)$, though, it is expressed more easily in terms of the spherical polar coordinates $(t,R,y^A)$ constructed from harmonic coordinates as follows:
\begin{subequations}
\label{eq:finite_transform}
\ba
\label{eq:R}
u &=&t-R -\frac{2M}{\lambda_0} \ln R\,,\\
r &=& R + BM - \frac{\psi}{2}\,,\\
\label{eq:y}
x^A &=& y^A\,.
\ea
\end{subequations}
The coordinates $(u,r,x^A)$ resemble, but are not precisely Bondi coordinates at the order in $1/r$ at which we work, because they do not enforce all of the required properties of the Bondi-gauge metric.
We will thus write this ``intermediate'' metric as $g_\mathrm{(I)}^{\mu\nu}$, and it can be computed from the harmonic-gauge metric using the transformation law for the components of a rank-two contravariant tensor:
\begin{equation}
    g_\mathrm{(I)}^{\mu\nu} = g_\mathrm{(H)}^{\alpha\beta} \frac{\partial x^\mu}{\partial X^\alpha} \frac{\partial x^\nu}{\partial X^\beta} \, .
\end{equation}
A somewhat lengthy, but otherwise straightforward calculation shows that this metric can be written in the form
\ba \label{eq:gmunuI}
g_\mathrm{(I)}^{\mu\nu} = \eta_\mathrm{(B)}^{\mu\nu} - \frac{1}{r}h_\mathrm{(I)}^{\mu\nu}\,,
\ea
where $\eta^\mathrm{(B)}_{\mu\nu}$ is the leading order part of the inverse Bondi metric in Eq.~\eqref{eq:InverseBondiMetric} when $\partial_u \Xi = \partial_u \lambda_1$.
The coefficients of the relative $1/r$ corrections to $\eta_\mathrm{(B)}^{\mu\nu}$ we denoted by $h_\mathrm{(I)}^{\mu\nu}$, and they are given by
\begin{subequations}
\label{eq:hImunu}
\ba
h_\mathrm{(I)}^{rr} &=& \frac{2M}{\lambda_0}(\partial_u\psi)^2 
-\left(\frac{\tilde H}{2}-2B M+\psi\right)\partial_u\psi \nonumber\\
&&+H_{ij}n^in^j+2B M+  O \lb r^{-1} \rb\,,\\
h_\mathrm{(I)}^{ur} &=& -\frac M{\lambda_0} \partial_u\psi 
+\frac{\Xi}{\lambda_0}+\frac{\tilde H}{2} + O \lb r^{-1} \rb\,,\\
r \, h_\mathrm{(I)}^{rA}&=& -\lb H_{0i}\eth^A n^i-\frac{1}{2}\eth^A\psi\rb + O \lb 
r^{-1} \rb\,,\\
h_{\mathrm(I)}^{uu} &=&  O \lb r^{-1}\rb\,,\\
h_\mathrm{(I)}^{uA} &=&  O \lb r^{-2} \rb\,,\\
r^2 h_\mathrm{(I)}^{A B} &=& H_{ij}\eth^{A} n^i\eth^{B} n^j+\psi q^{AB} + O \lb 
r^{-1} \rb\, . 
\label{eq:hI_AB}
\ea
\end{subequations}
To arrive at Eq.~\eqref{eq:hImunu}, we used the conditions in Eq.~\eqref{eq:HarmonicGauge} and we expressed $\partial_t\psi$ in terms of $\partial_u\psi$ (and other terms) using the derivatives of the first two lines in Eq.~\eqref{eq:finite_transform} and the chain rule:
\be
\partial_t \psi=\partial_u\psi -\frac{M}{r\lambda_0}(\partial_u\psi)^2 \, .
\ee
Note that nonlinear terms involving $\psi$ appear here and elsewhere in $h_{I}^{rr}$, because the coordinate transformations in~\eqref{eq:R}--\eqref{eq:y} involve $\psi$ at order $O(R^0)$. 

The metric is similar to a Bondi-Sachs form to the order in $1/r$ at which we work, in the sense that the Bondi gauge conditions $g_\mathrm{(I)}^{uu} = g_\mathrm{(I)}^{uA} = 0$ are satisfied at this order and the right-hand side of Eq.~\eqref{eq:hI_AB} is traceless with respect to $q_{AB}$ (thereby being consistent with the determinant condition of Bondi gauge).
Note, however, that the $ur$, $rr$, and $rA$ components of $h_{(I)}^{\mu\nu}$ do not satisfy the modified Einstein equations in Bondi-Sachs gauge, as they are not consistent with the form of the inverse metric in Eq.~\eqref{eq:InverseBondiMetric}.
The metric can be put into Bondi gauge with a perturbative (in $1/r$) coordinate transformation, as we describe next.

We will parametrize the perturbative coordinate transformation in terms of a vector $\xi^\mu$ which effects the coordinate transformation $x^\mu \rightarrow x^\mu + \xi^\mu$.
This coordinate transformation will take the part of the metric $h_\mathrm{(I)}^{\mu\nu}/r$ and bring it to the Bondi-Sachs form $h_\mathrm{(B)}^{\mu\nu}/r$, through the transformation
\begin{equation} \label{eq:Lie_xi_eta}
    \frac 1r h_\mathrm{(B)}^{\mu\nu} = \frac 1r h_\mathrm{(I)}^{\mu\nu} + \mathcal L_{\vec\xi} \eta_{(B)}^{\mu\nu} \, ,
\end{equation}
where $\mathcal L_{\vec\xi}$ is the Lie derivative along $\vec \xi$.
To solve for the perturbative gauge vector that generates this transformation, we first write the components of $\xi^\mu$ as
\begin{equation} \label{eq:xi_vec_r}
    \xi^\mu = \frac 1r (\xi^u_{(1)},\xi^r_{(1)},\xi^A_{(1)}/r) \, ,
\end{equation}
where the functions $\xi^\mu_{(1)}$ depend on $u$ and $x^A$ (not $r$).
These lead to a set of partial differential equations for the $\xi^\mu_{(1)}$ that can be integrated by requiring that $h_\mathrm{(I)}^{\mu\nu}$ be transformed into Bondi-Sachs form.
Before giving the full details of this procedure, it is worth noting that given the form of the components of the vector $\xi^\mu$ in Eq.~\eqref{eq:xi_vec_r}, the radiative data $\lambda_1$ and $c_{AB}$ can be related to harmonic-gauge data and the finite coordinate transformation~\eqref{eq:finite_transform} without needing the full expression for $\xi^\mu$ in~\eqref{eq:xi_vec_r}.

For the scalar field, because in both harmonic and Bondi gauges, the field has an expansion of the form $\lambda = \lambda_0 + O(r^{-1})$ [where $\lambda_0$ is constant and the coefficient of the $O(r^{-1})$ term is denoted $\Xi(t-R,y^A)$ in harmonic coordinates and $\lambda_1(u,x^A)$ in Bondi coordinates], then the transformation parametrized by $\xi^\mu$ in Eq.~\eqref{eq:xi_vec_r} will not change $\Xi$ or $\lambda_1$.
Thus, one must have that 
\begin{equation} \label{eq:lambda2Xi}
    \lambda_1(u,x^A) = \Xi(t-R,y^A) \, ,
\end{equation}
where $u$ is related to $t-R$ (and $x^A$ to $y^A$) by the transformations in Eq.~\eqref{eq:finite_transform}.
For $c_{AB}$, a direct calculation shows that $\mathcal L_{\vec\xi} \eta_{(B)}^{AB}$ is of order $r^{-4}$ for $\xi^\mu$ in~\eqref{eq:xi_vec_r}.
This implies that 
\begin{equation}
    c^{AB} = H_{ij} \eth^A n^i \eth^B n^j + \psi q^{AB} \, .
\end{equation}
Using the definition of $\tilde H_{\mu\nu}$ in Eq.~\eqref{eq:htildeToh}, the harmonic gauge conditions in~\eqref{eq:HarmonicGauge}, and the fact that $q^{AB} = \delta_{ij} \eth^A n^i \eth^B n^j$, this equation can be recast as
\begin{equation}
    c^{AB} = \tilde H_{ij} \eth^A n^i \eth^B n^j - \frac 12 \tilde H q^{AB} \, .
\end{equation}
After using Eq.~\eqref{eq:HarmonicGauge} again, it follows that $c^{AB}$ is related to just the transverse-traceless (TT) part of $\tilde H_{ij}$ as
\begin{equation} \label{eq:cAB2HijTT}
    c^{AB}(u,x^A) = \tilde H_{ij}^\mathrm{TT}(t-R,y^A) \eth^A n^i \eth^B n^j \, ,
\end{equation}
where again $u$ is related to $t-R$ and $x^A$ to $y^A$ by the transformations in~\eqref{eq:finite_transform}.

For our purposes of relating the radiative data in Bondi coordinates to that in harmonic coordinates, Eqs.~\eqref{eq:lambda2Xi} and~\eqref{eq:cAB2HijTT} provide the solution.
However, for completeness we do compute the form of the required gauge vector $\xi^\mu$ needed to complete the transformation from harmonic to Bondi coordinates.
The components of the gauge vector in Eq.~\eqref{eq:xi_vec_r} can be constrained from the $ur$, $rA$, and $rr$ components of Eq.~\eqref{eq:Lie_xi_eta}, which state, respectively,
\begin{subequations} \label{eq:xi_mu_pdes}
\begin{align}
    \partial_u \xi^u_{(1)} = {} & h_\mathrm{(B)}^{ru} - h_\mathrm{(I)}^{ru} \, , \\
    \partial_u \xi^A_{(1)} = {} & r(h_\mathrm{(B)}^{rA} - h_\mathrm{(I)}^{rA}) \, , \\
    2\partial_u \xi^r_{(1)} + \xi^u_{(1)} (\partial_u)^2 (\lambda_1/\lambda_0) = {} & h_\mathrm{(B)}^{rr} - h_\mathrm{(I)}^{rr} \, .
\end{align}
\end{subequations}
By substituting the relationships in Eqs.~\eqref{eq:lambda2Xi} and~\eqref{eq:cAB2HijTT} into Eq.~\eqref{eq:InverseBondiMetric}, extracting the relevant components of $h_\mathrm{(B)}^{\mu\nu}$, and using the expressions for $h_\mathrm{(I)}^{\mu\nu}$ in Eq.~\eqref{eq:hImunu}, it is straightforward to integrate the first two lines in Eq.~\eqref{eq:xi_mu_pdes} to obtain expressions for $\xi^u_{(1)}$ and $\xi^A_{(1)}$.
Once $\xi^u_{(1)}$ has been determined, integrating the final line of Eq.~\eqref{eq:xi_mu_pdes} to determine $\xi^r_{(1)}$ is also, in principle, straightforward.
There is one subtlety in this procedure: $h_\mathrm{(B)}^{rr}$ involves the Bondi mass aspect $\mathcal M$, which has not yet been determined from metric quantities in harmonic gauge.
Because the mass aspect satisfies the conservation equation~\cite{Tahura:2020vsa}
\begin{align}
    \partial_u \mathcal M = & - \frac 18 N_{AB} N^{AB} + \frac 14 \eth_A \eth_B N^{AB} \nonumber \\
    & -(3+2\omega) \frac 1{4\lambda_0^2} (\partial_u \lambda_1)^2 + \frac 1{4\lambda_0} \ETH^2 \partial_u \lambda_1 \, ,
\end{align}
it is possible to integrate this equation and express $\mathcal M$ in terms of harmonic-gauge quantities using Eqs.~\eqref{eq:lambda2Xi} and~\eqref{eq:cAB2HijTT}.
(To simplify the notation, however, we will not write this out in detail below, and we will write this quantity just as $\mathcal M$.)
The result of performing these integrations and using the harmonic gauge conditions in~\eqref{eq:HarmonicGauge} is that the components of the vector $\xi^\mu$ are given by
\begin{subequations}
\begin{align}
&\xi^u = -\frac{1}{2r}\int dt \tilde  H + \frac{M}{\lambda_0^2 r} \Xi \,, \\ 
&\xi^r = -\frac{1}{2r}\int dt \left[\frac{\xi^u_{(1)}}{\lambda_0}\partial_u^2\Xi + \frac{2M}{\lambda_0^3}(\partial_u \Xi)^2 -2\mathcal M + \tilde H_{00} \right. \nonumber\\
&\qquad\qquad\quad\left. -\frac{1}{2}\frac{\Xi}{\lambda_0^2}\partial_u\Xi -\left(\frac{1}{2}\tilde H-\frac{2M}{\lambda_0}\right)\left(1+\frac{\partial_u\Xi}{\lambda_0}\right) \right] \, ,\\
&\xi^A = -\frac{1}{r^2}\int dt \left[\frac{1}{2}\eth_B\left(\tilde H_{ij}^\mathrm{TT} \eth^A n^i \eth^B n^j\right) + H_{ij}n^i \eth^A n^j\right]\,.
\end{align}
\end{subequations}
This transformation, along with the finite transformation in Eq.~\eqref{eq:finite_transform}, brings the metric into the form in Eq.~\eqref{eq:InverseBondiMetric}, in which $\lambda_1$ and $c_{AB}$ are related to the harmonic-gauge quantities $\Xi$ and $\tilde H_{ij}^\mathrm{TT}$ by Eqs.~\eqref{eq:lambda2Xi} and~\eqref{eq:cAB2HijTT}.

\section{Estimates of ordinary memory effects in Brans-Dicke theory}\label{Sec:Ordinary_Memory}

\subsection{Ordinary displacement memory effect}

The contribution to the ordinary memory effect comes from the charge rather than the flux terms in Eq.~\eqref{eq:DispMemoryFlux}, i.e.:
\ba
\int \! d^2\Omega \, \alpha \,  \ETH^2 (\ETH^2+2) \Delta \Theta^\mathrm{O} = 8\Delta \! \int \! d^2\Omega \, \alpha\lb\mathcal M-\frac{1}{4\lambda_0}\ETH^2\lambda_1\rb\,.\nonumber\\
\ea
Expanding $\Delta\mathcal M$, $\Delta\Theta^\mathrm{O}$ and $\Delta \lambda_1$ in spherical harmonics, the moments $\Delta\Theta^\mathrm{O}$ are given by
\begin{equation}\label{eq:Ordinary_Thetalm}
\Delta \Theta_{lm}^\mathrm{O} =\frac{(l+2)!}{(l-2)!}\left[8\Delta \mathcal M_{lm}+\frac{2}{\lambda_0}l(l+1)\Delta \lambda_{1(lm)}\right] \,. 
\end{equation}
We would like to estimate if the quantity $\Delta \Theta_{lm}^\mathrm{O}$ is of a similar PN order to the nonlinear and null parts of the memory $\Delta \Theta_{lm}^\mathrm{T}$ and $\Delta \Theta_{lm}^\mathrm{S}$ that were computed in Sec.~\ref{subsec:displacementmemory} for any of the specific values of $l=2$, 4, or 6 and $m=0$.
To do so, we will focus on the moment $\Delta \Theta_{20}^\mathrm{O}$ for simplicity (the other three moments will have the same, or a higher PN order). 

One natural way to compute the ordinary memory would be to directly evaluate the moments $\Delta\lambda_{1(20)}$ and $\Delta\mathcal M_{20}$.
Using Eq.~\eqref{eq:scalar_field}, we can show that $\Delta\lambda_{1(20)}$ is at least $\mathcal O (x^2)$; thus, the scalar field's contribution to the memory effect is of a higher PN order than the Newtonian order at which we work.
We do not have an independent expression for $\Delta\mathcal M_{20}$ that would allow us to directly compute $\Delta \Theta_{20}^\mathrm{O}$ (although we already verified that the contribution from the $\Delta\lambda_{1(20)}$ is of a higher PN order). 
Instead, we can compute $\Delta\Theta_{20}^\mathrm{O}$ directly from the waveform that were already computed in Eqs.~(7.1) and~(7.2a) of~\cite{Lang:2013fna} to verify that $\Delta\mathcal M_{20}$ would not contribute at Newtonian order.
Specifically, we contract the Newtonian-order expression with the polarization tensors $e_+^{ij} - i e_\times^{ij}$, multiply by the spin-weighted spherical harmonic ${}_{-2}\bar Y_{20}$ to obtain $U_{20}$ and then rescale it to obtain $\Delta \Theta_{20}^\mathrm{O}$.
We find that the Newtonian-order result vanishes, and there is thus no Newtonian-order ordinary displacement memory.

\subsection{Ordinary spin memory effect}

The ordinary part of the spin memory effect can be computed from  Eq.~\eqref{eq:BalanceSpinMemory} with just the term~\eqref{eq:QSpin} on the right-hand side:
\begin{align}
  \int \! d^2\Omega\,\gamma\ETH^2(\ETH^2 + 2) \Delta \Sigma^\mathrm{0} & = -8\Delta \! \int \! d^2\Omega \, \epsilon^{AD}\eth_D\gamma\left[- 3L_A\right.\nonumber\\
& - \left.\frac 1{4\lambda_0}(c_{AB} \eth^B \lambda_1 - \lambda_1 \eth^B c_{AB})\right]\,.
\end{align}
The terms $c_{AB} \eth^B \lambda_1 - \lambda_1 \eth^B c_{AB}$ on second line of the equation will not contribute at Newtonian order for the spin memory effect (i.e, at order $x^{-1/2}$), because both $c_{AB}$ and $\lambda_1$ involve non-negative powers of $x$ in the PN expansion, so their product will not be a negative power of $x$. 
The only term that could contribute to the spin memory effect comes from the change in the angular momentum aspect, $L_A$. 

We do not have an expression for $L_A$ in terms of harmonic-gauge metric functions, which (analogously to the case of the mass aspect and ordinary displacement memory effect) prohibits a direct calculation of the ordinary spin memory effect.
In addition, it is not possible to directly check the time integral of the waveform, because the Newtonian-order terms in the spin memory effect arise from formally 2.5PN-order terms in the waveform that are then integrated with respect to retarded time; however, the waveform has only been computed to 2PN order in~\cite{Lang:2013fna}. 
While we cannot then be certain that the ordinary memory terms do not contribute at the same order because of additional nonlinear terms in the near zone, we can estimate the size of the effect in linearized theory.

The ordinary spin memory effect would arise at the lowest PN order from changes in $\Delta\Sigma_{30}^\mathrm{O}$, which is proportional to the retarded time integral of the radiative moment $V_{30}$. 
Because at leading PN order, $V_{30}$ is related to three time derivatives of the source current octopole $J_{30}$, then $\Delta \Sigma_{30}^\mathrm{O}$ should be proportional to $\ddot J_{30}$. 
By dimensional analysis, $J_{30}$ is proportional to $M v a^3$ (or see, e.g.,~\cite{Blanchet:2013haa}); thus, $\ddot J_{30}$ scales as $M v a \dot a^2$. 
This scales with the PN parameter as $\xi x^{9/2} + x^{11/2}$, which would be a 6PN correction to the nonlinear and null effects.
We thus anticipate from these arguments in linearized theory that the ordinary part of the spin memory will be small.

\bibliography{memory}

\end{document}